\documentclass[aps,prd,groupedaddress,nofootinbib,amssymb,eqsecnum,epsfig]{revtex4}
\usepackage{graphicx}
\usepackage{bm}
\usepackage{amsmath}
\usepackage{color}
\usepackage{amsfonts}

\def \lleq {\lower0.9ex\hbox{ $\buildrel < \over \sim$} ~}
\def \ggeq {\lower0.9ex\hbox{ $\buildrel > \over \sim$} ~}

\def \lcdm    {$\Lambda$CDM }

\def \beq  {\begin{equation}}
\def \eeq  {\end{equation}}
\def \ber  {\begin{eqnarray}}
\def \eer  {\end{eqnarray}}
\def \lcdm  {$\Lambda$CDM }

\newcommand{\magcir}{\raise
-3.truept\hbox{\rlap{\hbox{$\sim$}}\raise4.truept\hbox{$>$}\ }}
\newcommand{\mincir}{\raise
-3.truept\hbox{\rlap{\hbox{$\sim$}}\raise4.truept\hbox{$<$}\ }}

\begin{document}
\newcommand{\newc}{\newcommand}

\newcommand{\ben}{\begin{eqnarray}}
\newcommand{\een}{\end{eqnarray}}
\newc{\be}{\begin{equation}}
\newc{\ee}{\end{equation}}
\newc{\ba}{\begin{eqnarray}}
\newc{\ea}{\end{eqnarray}}
\newc{\bea}{\begin{eqnarray*}}
\newc{\eea}{\end{eqnarray*}}
\newc{\D}{\partial}
\newc{\ie}{{\it i.e.} }
\newc{\eg}{{\it e.g.} }
\newc{\etc}{{\it etc.} }
\newc{\etal}{{\it et al.}}
\newcommand{\nn}{\nonumber}
\newc{\ra}{\rightarrow}
\newc{\lra}{\leftrightarrow}
\newc{\lsim}{\buildrel{<}\over{\sim}}
\newc{\gsim}{\buildrel{>}\over{\sim}}

\title{Cosmological perturbations and observational constraints on nonlocal massive gravity}

\author{Savvas Nesseris$^{1}$}

\author{Shinji Tsujikawa$^{2}$}

\email{savvas.nesseris@uam.es, shinji@rs.kagu.tus.ac.jp}

\affiliation{$^1$ Instituto de F\'isica Te\'orica UAM-CSIC, Universidad Auton\'oma de Madrid,
Cantoblanco, 28049 Madrid, Spain\\
$^2$ Department of Physics, Faculty of Science, Tokyo University of Science, 1-3, Kagurazaka,
Shinjuku-ku, Tokyo 162-8601, Japan}

\date{\today}

\begin{abstract}

Nonlocal massive gravity can provide an interesting explanation for the late-time cosmic acceleration, with a dark energy equation of state $w_{\rm DE}$ smaller than $-1$ in the past. We derive the equations of linear cosmological perturbations to confront such models with the observations of large-scale structures. The effective gravitational coupling to nonrelativistic matter associated with galaxy clusterings is close to Newton's gravitational constant $G$
for a mass scale $m$ slightly smaller than today's Hubble parameter $H_0$.
Taking into account the background expansion history as well as the evolution
of matter perturbations $\delta_m$, we test for these models with Type Ia Supernovae (SnIa) from  Union 2.1, the cosmic microwave background (CMB) measurements from Planck, a collection of baryon acoustic oscillations (BAO), and the growth rate data of $\delta_m$. Using a higher value of $H_0$ derived from its direct measurement ($H_0 \gtrsim 70$ km s$^{-1}$ Mpc$^{-1}$) the data strongly support the nonlocal massive gravity model ($-1.1 \lesssim w_{\rm DE} \lesssim -1.04$ in the past) over the $\Lambda$CDM model ($w_{\rm DE}=-1$), whereas for a lower prior (67 km s$^{-1}$ Mpc$^{-1}$ $\lesssim$ $H_0 \lesssim 70$ km s$^{-1}$ Mpc$^{-1}$) the two models are statistically comparable.

\end{abstract}


\maketitle

\section{Introduction}

Modified gravitational theories have received much attention in connection to the
dark energy problem \cite{review}.
In particular, the recent observational constraints
derived from Planck and other data show that the dark energy equation
of state $w_{\rm DE}$ smaller than $-1$ is favored \cite{Ade:2013zuv}.
This may imply the infrared modification of gravity from general relativity (GR),
because the models in the framework of GR --- such as quintessence \cite{quin} and
k-essence \cite{kes} --- generally predict $w_{\rm DE}$ larger than $-1$.

So far many dark energy models based on the large-distance modification
of gravity have been proposed --- including the Dvali-Gabadadze-Porrati (DGP)
model \cite{DGP}, $f(R)$ gravity \cite{fR}, Brans-Dicke
theories \cite{Brans}, and Galileons \cite{Galileons}.
In the DGP model the cosmic acceleration can be realized by the
gravitational leakage to the fifth dimension, but it suffers from
the incompatibility with observations \cite{DGPcon}
as well as the ghost problem \cite{Luty}.
In $f(R)$ gravity and Brans-Dicke theories it is possible to construct
viable dark energy models at the expense of designing scalar
potentials to be compatible with both cosmological and local gravity
constraints \cite{fR2}. In covariant Galileons there exists a tracker solution along
which $w_{\rm DE}$ evolves from $-2$ (matter era) to $-1$
(de Sitter era) \cite{DT12}, but only the late-time tracking solutions
are allowed from the joint data analysis of SnIa, CMB, and BAO \cite{Nesseris}.

Besides the theories mentioned above, massive gravity has recently received
significant attention due to the possibility of the late-time cosmic acceleration
with a mass scale $m$ of the order of today's Hubble parameter $H_0$.
In the original Fierz-Pauli theory \cite{fierzpauli} there exists a so-called van
Dam-Veltman-Zakharov (vDVZ) discontinuity \cite{DVZ} with which the
linearized GR cannot be recovered in the $m \to 0$ limit.
In the presence of nonlinear interactions the problem of the vDVZ
discontinuity can be cured \cite{Vainshtein}, but an instability mode
called the Boulware-Deser (BD) ghost
appears due to nonlinearities \cite{BDghost}.

De Rham, Gabadadze and Tolley (dRGT) constructed a massive gravity
theory \cite{deRham:2010kj} in which the BD ghost is absent.
In addition to acausality of the theory \cite{aca} and the requirement
of an external reference metric, there are some
problems when the dRGT theory is applied to the cosmology.
On the homogenous and isotropic cosmological background,
it was shown that at least one ghost exists among five propagating
degrees of freedom \cite{DGM}. The self-accelerating
solutions in the dRGT theory are also unstable against
scalar and vector perturbations \cite{ins}.
The possible way out of these problems is to break the
homogeneity or the isotropy of the Universe \cite{homoge,ani}
or to introduce other degrees of freedom \cite{quasi1,quasi2,quasi3}.

An alternative approach to massive gravity was recently suggested by
Jaccard {\it et al.} \cite{Jaccard}, who introduced nonlocal terms
to obtain fully covariant equations of motion without referring to
any external reference metric (see also Refs.~\cite{DHK,Deser,Jhingan,Koivisto,Denonlocal,Zhang,Elizalde,Park,Deser13}
for related works). This theory --- dubbed non-local massive gravity (NLMG)--- respects
causality and reduces to a massless one without the vDVZ discontinuity in the
$m \to 0$ limit. The covariant equations of motion are given by
\be
G_{\mu\nu}-m^2 (\square_{\rm ret}^{-1}G_{\mu\nu})^{\rm T}
=8\pi G\,T_{\mu\nu}\,,
\label{nonlocal0}
\ee
where $G_{\mu \nu}$ is the Einstein tensor, $T_{\mu \nu}$ is
the energy-momentum tensor, $G$ is the gravitational constant,
$\square_{\rm ret}^{-1}$ is the inverse of d'Alembertian operator
computed with the retarded Green's function,
and the superscript T represents the extraction of the transverse part.

The background cosmological dynamics based on Eq.~(\ref{nonlocal0})
was studied in Ref.~\cite{Modesto:2013jea}.
There is a rapidly growing scalar mode responsible for the late-time
cosmic acceleration, in which case
the dark energy equation of state evolves from
$w_{\rm DE}=-1.725$ (matter era) to
$w_{\rm DE}=-1.506$ (accelerated era).
Since the Planck data combined with the SnIa and WMAP polarization
data placed the bound $w_{\rm DE}=-1.13^{+0.13}_{-0.14}$ (95\,\%\,C.L.)
for constant $w_{\rm DE}$ \cite{Ade:2013zuv}, the NLMG model (\ref{nonlocal0}) is
in tension with the current observations of CMB and SnIa.
In order to avoid the rapid growth of the scalar mode, we also require
that the mass $m$ is much smaller than $H_0$.

Alternatively, Maggiore \cite{Maggiore:2013mea} proposed a model
given by the field equation
\be
G_{\mu\nu}-\frac13 m^2 (g_{\mu \nu}\square_{\rm ret}^{-1}R)^{\rm T}
=8\pi G\,T_{\mu\nu}\,,
\label{nonlocalmo}
\ee
where $g_{\mu \nu}$ is the metric tensor and $R$ is the Ricci scalar.
In this case the strong instability of a scalar mode present in the theory
(\ref{nonlocal0}) is avoided, so that the dark energy equation of state
does not significantly deviate from $-1$
($w_{\rm DE} \approx -1.1$ in the deep matter era).
The model has a predictive power due to the presence
of a single parameter $m$ alone.
For today's dark energy density parameter
$\Omega_{\rm DE}^{(0)} \simeq 0.68$, the mass $m$ is fixed to be
$m \simeq 0.67 H_0$ \cite{Maggiore:2013mea,Foffacosmo}.
It was also shown that the general relativistic behavior can
be recovered inside the solar system \cite{Keha}.

The retarded operator $\square_{\rm ret}^{-1}$ mentioned above is required
for causality. However, the variation of some nonlocal action involving the inverse
d'Alembertian operator $\square^{-1}$ (such as $\phi\, \square^{-1} \phi$ with $\phi$ being
a scalar) usually symmetrizes the Green's function \cite{Foffa:2013sma}.
Hence the field equations of motion involving the retarded operator $\square_{\rm ret}^{-1}$
do not follow from a variational principle from some nonlocal action.
Such a retarded nonlocal operator will only emerge when some classical or quantum
averaging prescription is performed in a more fundamental local quantum field theory (QFT).
In this sense, the field equations of motion (\ref{nonlocal0}) and (\ref{nonlocalmo})
should be considered only as effective classical equations that do not have a direct link
to the action of a nonlocal QFT \cite{Foffa:2013sma}. This is also the approach we follow in our paper.

Furthermore, if we use a quadratic action of gravitational waves associated with
the perturbation equation of the theory (\ref{nonlocalmo}) by simply replacing
$\square_{\rm ret}^{-1}$ with $\square^{-1}$, the resulting
propagator apparently involves a ghostlike massive scalar \cite{Maggiore:2013mea}.
Foffa {\it et al.} \cite{Foffa:2013sma} showed that this apparent ghost is not
a propagating degree of freedom and in the $m \to 0$ limit it smoothly approaches
a nonradiative degree of freedom of GR. This implies that we should
regard Eq.~(\ref{nonlocalmo}) as an effective classical equation of motion rather
than promoting it directly to a full QFT (which typically involves some classical or
quantum averaging). The issue of quantization---including ghosts---would
be addressed in an underlying fundamental theory with a possible ultraviolet completion.

In this paper we study the cosmology (at the classical level) and observational constraints on the NLMG models. In Sec.~\ref{backsec} the background equations of motion are derived for general models including (\ref{nonlocal0}) and (\ref{nonlocalmo}).
We then discuss the evolution of $w_{\rm DE}$ as well as the mass scale $m$ constrained from the background cosmology. In Sec.~\ref{persec} we obtain the full equations of linear cosmological perturbations for the NLMG model (\ref{nonlocalmo}). We also discuss the behavior of perturbations for the subhorizon modes relevant to large-scale structures. In Sec.~\ref{resultscomp} we confront the NLMG model (\ref{nonlocalmo})
with the latest observations of SnIa, CMB, BAO, and redshift-space distortions.
Section ~\ref{consec} is devoted to the conclusions.

\section{Background equations of motion}
\label{backsec}

We start with the following equations of motion of the NLMG
models\footnote{Following Ref.~\cite{Jaccard} we use
the metric signature $(-,+,+,+)$ in this paper.
Note that the metric signature used in Ref.~\cite{Modesto:2013jea}
is $(+,-,-,-)$.}
\be
G_{\mu\nu}-m^2 \square_{\rm ret}^{-1} \left( a_1 R_{\mu\nu}
+a_2\,g_{\mu\nu} R \right)^{\rm T}=8\pi G \, T_{\mu\nu}\,,
\label{nonlocal1}
\ee
where $a_1$ and $a_2$ are constants, $R_{\mu \nu}$ is the Ricci tensor,
$\square_{\rm ret}^{-1}$ is the inverse of the d'Alembertian operator
computed by using the retarded Green's function due to causality \cite{Foffacosmo}.
The model (\ref{nonlocal0}) corresponds to $a_1=1$ and $a_2=-1/2$,
whereas the model (\ref{nonlocalmo}) is characterized by
$a_1=0$ and $a_2=1/3$.

We now introduce a tensor $S_{\mu \nu}$ obeying the relation
\be
\Box S_{\mu \nu} =a_1 R_{\mu\nu}
+a_2\,g_{\mu\nu} R \,.
\label{Smure}
\ee
In order to respect the continuity equation
$\nabla^{\mu} T_{\mu \nu}=0$ of matter in Eq.~(\ref{nonlocal1}),
we pick up the transverse part $S_{\mu \nu}^{\rm T}$ of
the symmetric tensor $S_{\mu \nu}$ satisfying
$\nabla^{\mu} S_{\mu \nu}^{\rm T}=0$, that is
\be
G_{\mu\nu}-m^2S_{\mu \nu}^{\rm T} =8\pi G \, T_{\mu\nu}\,.
\label{nonlocal2}
\ee
The tensor $S_{\mu \nu}$ can be decomposed as \cite{Porrati}
\be
S_{\mu \nu}=S_{\mu \nu}^{\rm T}
+(\nabla_{\mu} S_{\nu}
+\nabla_{\nu} S_{\mu})/2\,.
\label{Smunu}
\ee

Let us consider the flat Friedmann-Lema\^{i}tre-Robertson-Walker
(FLRW) spacetime described by the metric
$ds^2=-dt^2+a^2(t)\delta_{ij}dx^i dx^j$, where $t$ is the cosmic time.
On this background the vector $S_{\mu}$ has a time component
$S_0$ alone, so that
\be
(S^0_0)^{\rm T}=u+\dot{S}_0\,,\qquad
(S^i_i)^{\rm T}=v+3HS_0\,,
\ee
where $u \equiv S^0_0$ and $v \equiv S^i_i$,
$H \equiv \dot{a}/a$, and
a dot represents a derivative with respect to $t$.
For the energy-momentum tensor $T_{\mu \nu}$,
we take into account a perfect fluid obeying the
continuity equation
\be
\dot{\rho}+3H (\rho+P)=0\,,
\label{coneq}
\ee
where $\rho$ is the energy density and
$P$ is the pressure of the fluid.
{}From Eq.~(\ref{nonlocal1}) we obtain the following
equations of motion
\ba
& & 3H^2+m^2 (u+\dot{S}_0)=8\pi G \rho\,,
\label{Heq}\\
& & 2\dot{H}+3H^2+\frac{m^2}{3}
(v+3HS_0)=-8\pi G P\,.\label{Heq2}
\ea
{}From the (00) and $(ii)$ components of Eq.~(\ref{Smure})
it follows that
\ba
& & \ddot{u}+3H \dot{u}-6H^2u+2H^2 v
=-3(a_1+4a_2)H^2-3(a_1+2a_2)\dot{H}\,,\label{ueq}\\
& & \ddot{v}+3H \dot{v}-2H^2v+6H^2u
=-9(a_1+4a_2)H^2-3(a_1+6a_2)\dot{H}\,.\label{veq}
\ea
The divergence of Eq.~(\ref{Smunu}) gives
$2\nabla^{\mu} S_{\mu \nu}=\nabla^{\mu}
(\nabla_{\mu} S_{\nu}+\nabla_{\nu} S_{\mu})$.
The $\nu=0$ component of this equation reads
\be
\ddot{S}_0+3H \dot{S}_0-3H^2 S_0
=-\left( \dot{u}+3H u-H v \right)\,.
\label{Seq}
\ee

In order to simplify the analysis, we define
\be
U \equiv u+v\,,\qquad V \equiv u-v/3\,,
\label{Udef}
\ee
by which $u=(U+3V)/4$ and $v=(3/4)(U-V)$.
The field $U$ corresponds to the trace part of the tensor
$S^{\mu}_{\nu}$, whereas the field $V$ characterizes
the difference between the time and spatial
diagonal components of $S^{\mu}_{\nu}$.
On the right-hand side of Eqs.~(\ref{Heq}) and (\ref{Heq2})
we take into account the contribution of nonrelativistic matter
(density $\rho_m$, pressure $P_m=0$) and radiation
(density $\rho_r$, pressure $P_r=\rho_r/3$).
We can write Eqs.~(\ref{Heq}) and (\ref{Heq2})
in the following forms
\ba
&& 3H^2=8\pi G (\rho_m+
\rho_r+\rho_{\rm DE} )\,,\label{Fri1}\\
& &2\dot{H}+3H^2=-8\pi G (P_r+P_{\rm DE})\,,
\label{Fri2}
\ea
where
\be
\rho_{\rm DE}=\frac{m^2}{32\pi G} (4\zeta X-4X'-U-3V)\,,\qquad
P_{\rm DE}=\frac{m^2}{32\pi G} (U-V+4X)\,,
\label{rhode}
\ee
and
\be
X=HS_0\,,\qquad
\zeta=\frac{H'}{H}\,.
\ee
A prime represents a derivative with respect to $N=\ln (a/a_i)$,
where $a_i$ is the initial scale factor.
The dark energy equation of state is then given by
\be
w_{\rm DE}=\frac{P_{\rm DE}}{\rho_{\rm DE}}
=\frac{U-V+4X}{4\zeta X-4X'-U-3V}\,,
\label{wdere}
\ee
whereas the effective equation of state is
$w_{\rm eff}=-1-2\zeta/3$.
{}From Eqs.~(\ref{ueq})--(\ref{Seq}) it follows that
\ba
& &U''+(3+\zeta)U'=-6(a_1+4a_2)(2+\zeta)\,,\label{auto1}\\
& &V''+(3+\zeta)V'-8V=-2a_1 \zeta\,,\label{auto2}\\
& & X''+(3-\zeta)X'-(3+3\zeta+\zeta')X
=-\frac14
\left( U'+3V'+12V \right)\,.\label{auto3}
\ea
{}From Eq.~(\ref{Fri2}) the functions $\zeta$
and $\zeta'$ obey
\ba
\zeta &=& -\frac32-\frac{m^2}{8H^2} (U-V+4X)-\frac12 \Omega_r\,,
\label{zeta} \\
\zeta' &=& 2\Omega_r-3\zeta-2\zeta^2
-\frac{m^2}{8H^2} (U'-V'+4X')\,,
\label{zetad}
\ea
where we used the fact that the radiation density parameter
$\Omega_r=8\pi G \rho_r/(3H^2)$ satisfies
\be
\Omega_r'=-(4+2\zeta)\Omega_r\,.
\label{auto6}
\ee
{}From Eq.~(\ref{Fri1}) the matter density parameter
$\Omega_m=8\pi G \rho_m/(3H^2)$ is known to be
\be
\Omega_m=1-\Omega_r-\Omega_{\rm DE}\,,\quad
{\rm where } \quad
\Omega_{\rm DE}
=\frac{m^2}{12H^2}(4\zeta X-4X'-U-3V)\,.
\label{Omegam}
\ee
\begin{figure*}[t!]
\centering
\vspace{0cm}\rotatebox{0}{\vspace{0cm}\hspace{0cm}\resizebox{0.48\textwidth}{!}{\includegraphics{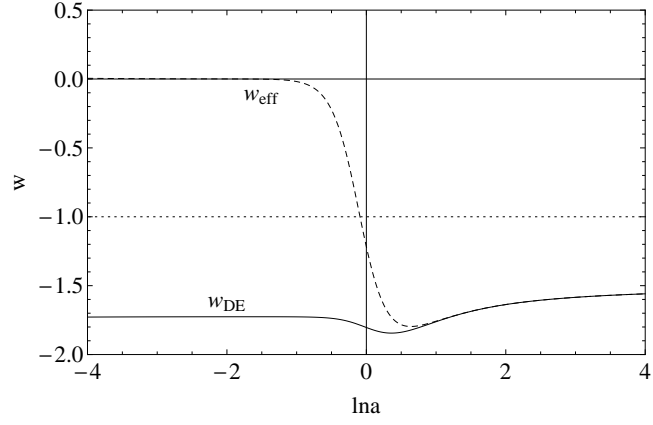}}}
\caption{Evolution of the dark energy equation of state $w_{\rm DE}$
and the effective equation of state $w_{\rm eff}$ versus $\ln a$ for the theory
with $a_1=0.01$ and $a_2=1/3$.
The present epoch corresponds to $\ln a=0$ (i.e., $a=1$).
\label{fig1}}
\end{figure*}
\begin{figure*}[t!]
\centering
\vspace{0cm}\rotatebox{0}{\vspace{0cm}\hspace{0cm}\resizebox{0.48\textwidth}{!}{\includegraphics{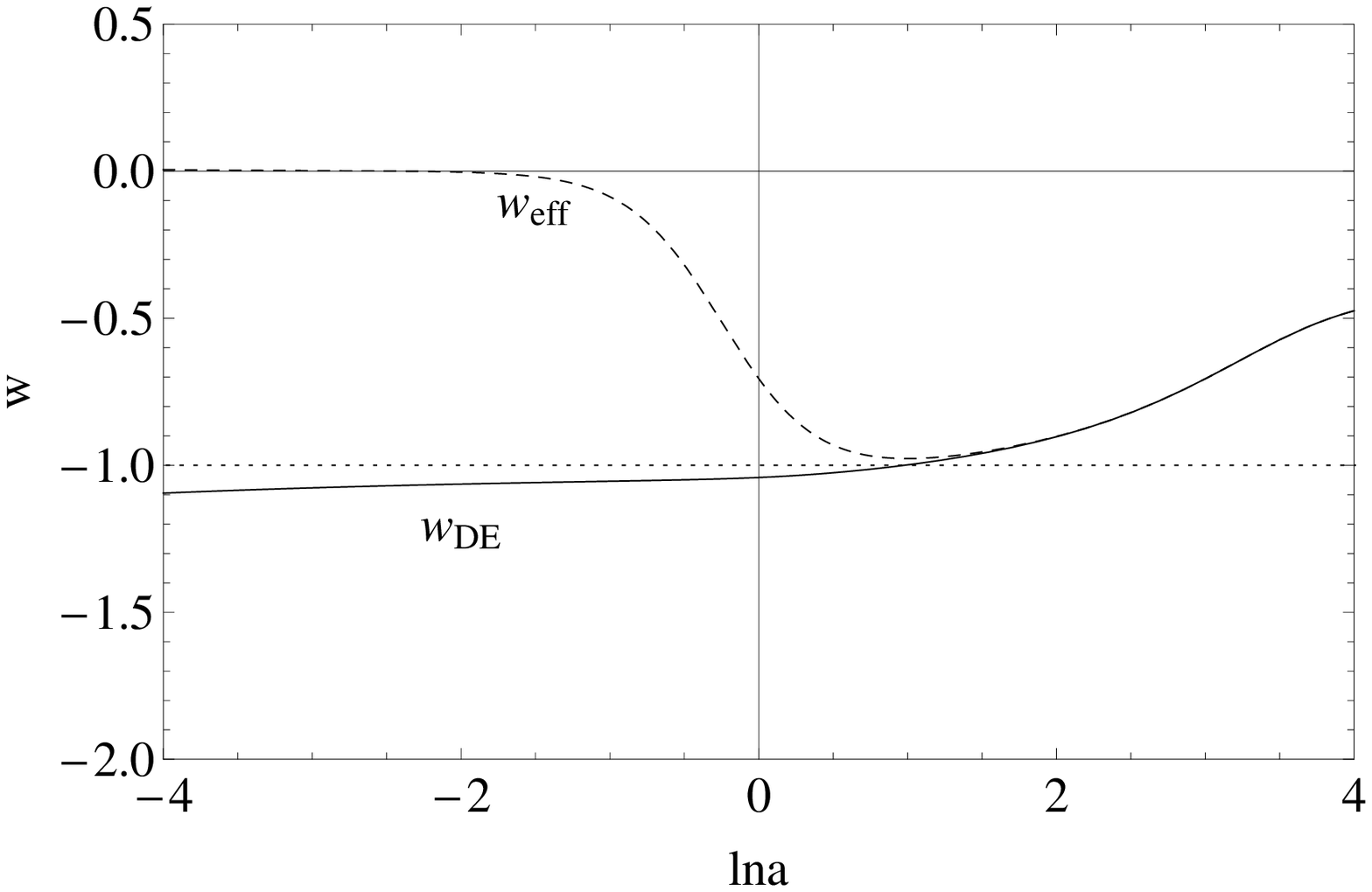}}}
\vspace{0cm}\rotatebox{0}{\vspace{0cm}\hspace{0cm}\resizebox{0.48\textwidth}{!}{\includegraphics{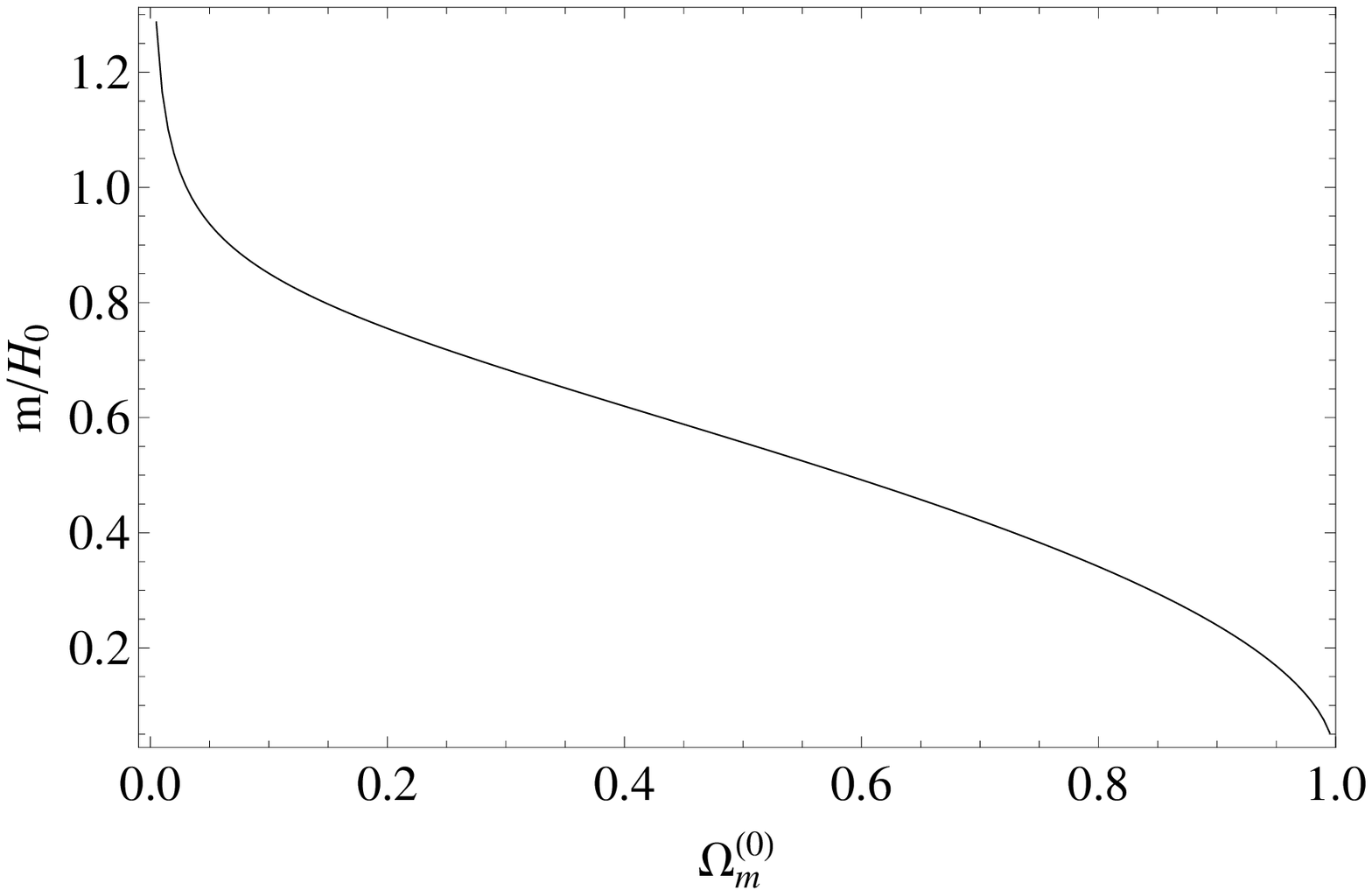}}}
\caption{Left: Evolution of the dark energy equation of state $w_{\rm DE}$
and the effective equation of state $w_{\rm eff}$ versus $\ln a$ for the theory
with $a_1=0$ and $a_2=1/3$.
Right: The mass scale $m$ (divided by $H_0$) versus today's
matter density parameter $\Omega_m^{(0)}$.
\label{fig2}}
\end{figure*}
The density parameter of radiation today
(corresponding to $a=1$) is given by
\be
\Omega_r^{(0)}=\Omega_{\gamma}^{(0)}
(1+0.2271 N_{\rm eff})\,,
\label{Omer}
\ee
where $\Omega_{\gamma}^{(0)}$ is the photon density parameter
and $N_{\rm eff}$ is the relativistic degrees of freedom.
We adopt the standard values $\Omega_{\gamma}^{(0)}
=2.469 \times 10^{-5}\,h^{-2}$
and $N_{\rm eff}=3.04$, where
$H_0=100\,h$ km s$^{-1}$ Mpc$^{-1}$ \cite{Hinshaw:2012fq}.
So, given that $\Omega_r^{(0)}$ is fixed from the CMB,
the only free parameter is $\Omega_m^{(0)}$.
{}From Eq.~(\ref{Omegam}) the mass ratio $m/H_0$
is calculated as
\be
\frac{m}{H_0}=\sqrt{\frac{12(1-\Omega_m-\Omega_r)}
{4\zeta X-4X'-U-3V}} \Biggl|_{a=1}\,.\label{mH0}
\ee

We integrate the dynamical equations of motion
(\ref{auto1})-(\ref{auto6}) with the initial conditions
\be
U(t_i)=\dot{U}(t_i)=V(t_i)=\dot{V}(t_i)=
X(t_i)=\dot{X}(t_i)=0\,,
\label{initial}
\ee
where $t_i$ corresponds to the time at the deep radiation era.
The initial conditions can be fixed as above once
the definition of the retarded inverse d'Alembertian
$\square_{\rm ret}^{-1}$ based on the retarded Green function is given.
Note that this issue was discussed in detail in Ref.~\cite{Foffacosmo}.
The initial time $t_i$ (taken at the deep radiation era) is regarded at
the epoch when the effective description of our classical equations of
motion becomes appropriate. The initial conditions (\ref{initial})
correspond to the minimal case in which the influence of the nonlocal
corrections to general relativity is negligible at $t=t_i$.
For example, the field $U$ is proportional to
$\square_{\rm ret}^{-1} R=-\int_{t_i}^t dt'a^{-3} (t') \int_{t_i}^{t'}
dt'' a^3 (t'')R(t'')$ and hence
$U(t_i)=\dot{U}(t_i)=0$ \cite{Foffacosmo}.
For given $m$, we numerically solve the background equations
of motion (\ref{auto1})--(\ref{auto6}) and evaluate (\ref{mH0}) to check
the consistency of the solutions.

{}From Eq.~(\ref{auto2}) it is clear that the field $V$ is unstable
for $a_1 \neq 0$ due to the presence of the $-8V$ term in Eq.~(\ref{auto2}) and the term $-2 a_1 \zeta$ on the right-hand side that acts as a source term if $a_1$ is not equal to zero. Even if all the fields $U$, $V$, $X$ and their first time derivatives are 0, we cannot avoid this instability. However, for $a_1=0$, the field $V$ is decoupled from the dynamical system because it can stay at $V=0$. In other words, for the equation of motion (\ref{nonlocalmo}) corresponding to the case $a_1=0$, the degree of freedom associated with the field $V$ does not appear. In fact, for the theory with $a_1=1$ and $a_2=-1/2$ \cite{Jaccard}, this instability leads to the rapid growth of $V$ and $X$,
by which the dark energy equation of state evolves as
$w_{\rm DE} \simeq -1.791$ (radiation era),
$w_{\rm DE} \simeq -1.725$ (matter era),
and $w_{\rm DE} \simeq -1.506$ (accelerated era) \cite{Modesto:2013jea}.
Unless $a_1$ is very close to 0, the same property also holds
for the theories with $a_1 \neq 0$.

In Fig.~\ref{fig1} we plot the evolution of $w_{\rm DE}$ and
$w_{\rm eff}$ for $a_1=0.01$ and $a_2=1/3$ with today's
matter density parameter $\Omega_m^{(0)}=0.323$.
As in the theory with $a_1=1$ and $a_2=-1/2$, the dark energy
equation of state during the matter-dominated epoch is around
$w_{\rm DE} \simeq -1.725$ and it finally approaches
the value $-1.506$. In this case the mass $m$ is much smaller than
$H_0$, which is required to avoid the early dominance of
dark energy \cite{Modesto:2013jea}.
The evolution of $w_{\rm DE}$ shown in Fig.~\ref{fig1}
is incompatible with the joint data analysis of CMB, SnIa,
and BAO, due to the large deviation from $-1$.
This conclusion holds not only from the WMAP data combined
with the SnIa and BAO measurements \cite{Nesseris,extended} but also from
the Planck data combined with the SnIa and WMAP polarization
measurements \cite{Ade:2013zuv}.

When $a_1=0$ the rhs of Eq.~(\ref{auto2}) vanishes, so that
the field $V$ does not contribute to the cosmological dynamics
for the initial conditions given in (\ref{initial}).
In the deep radiation era ($\zeta \simeq -2$ and
$\zeta' \simeq 0$) the field $U$ is almost frozen,
but it starts to evolve once $\zeta$ deviates from $-2$.
During the matter-dominated epoch ($\zeta \simeq -3/2$ and
$\zeta' \simeq 0$), integrations of Eqs.~(\ref{auto1})
and (\ref{auto3}) read
\ba
U &\simeq& -8a_2 N+c_1\,,\\
X &\simeq&  \frac43 a_2+c_2 e^{-(9-\sqrt{57})N/4}\,,
\label{Xint}
\ea
where $c_1$ and $c_2$ are integration constants, and
we neglected the decaying-mode solutions.
If the field value $|U|$ at the radiation-matter equality (identified as
$N=0$) is much smaller than 1, it follows that $|c_1| \ll 1$.
On using the solution (\ref{Xint}), the dark energy density
(\ref{rhode}) in the matter era is given by
\be
\rho_{\rm DE} \simeq \frac{m^2}{4\pi G} a_2
\left( N-1-\frac{c_1}{8a_2} \right)\,,
\ee
where the term $c_1/(8a_2)$ is much smaller than 1 for $|a_2|=O(1)$.
In the regime $N \gtrsim 1$, $\rho_{\rm DE}$ is positive for
\be
a_2>0\,.
\ee
This condition is assumed in the following discussion.

Neglecting the second term on the rhs of Eq.~(\ref{Xint}),
the dark energy equation of state (\ref{wdere}) in the deep
matter era reads
\be
w_{\rm DE} \simeq -\frac{1-16r}{1-24r}\,,\qquad
r \equiv -\frac{a_2}{3U}\,.
\label{wdeana}
\ee
In the regime $N \gtrsim 1$ we have $U \simeq -8a_2 N$ and
$r \simeq 1/(24N)$ for $a_2=O(1)$, so that
$w_{\rm DE} \simeq -[1-2/(3N)]/(1-1/N)$.
Hence the dark energy equation of state (\ref{wdeana})
is in the range $-1.11<w_{\rm DE}<-1$ for $N>4$.
At late times ($N \gg 1$) the terms $m^2/H^2$ grow in
Eqs.~(\ref{zeta}) and (\ref{zetad}), so
Eq.~(\ref{wdeana}) starts to lose its validity.

In the left panel of Fig.~\ref{fig2} the evolution of $w_{\rm DE}$ and $w_{\rm eff}$
is plotted for $a_1=0$, $a_2=1/3$, and $\Omega_m^{(0)}=0.323$.
The value $a_2=1/3$ was chosen to match with
the one in Ref.~\cite{Maggiore:2013mea},
but its precise value does not matter provided $a_2=O(1)$.
As estimated above, the dark energy equation of state exhibits mild
growth around $-1.1<w_{\rm DE} <-1.04$ by today
($-4<\ln a<0$). This mild variation of $w_{\rm DE}$ is followed
by more rapid growth toward $w_{\rm DE} \approx -0.5$ in the future.
In this case the mass is found to be $m/H_0 \simeq 0.67$,
which shows good agreement with the one derived in Ref.~\cite{Maggiore:2013mea}.
In the right panel of Fig.~\ref{fig2} we also plot the mass ratio $m/H_0$ versus
$\Omega_m^{(0)}$ in the range $\Omega_m^{(0)} \in [0, 1]$.
For larger $\Omega_m^{(0)}$ the mass $m$ gets smaller,
but it is typically of the order of $m/H_0=O(0.1)$.

\section{Cosmological perturbations}
\label{persec}

In this section we shall derive the equations of motion for linear cosmological
perturbations for the NLMG model (\ref{nonlocal1}) with $a_1=0$.
{}From Eq.~(\ref{Smure}) we have
$\square S^{\mu}_{\nu}=a_2 \delta^{\mu}_{\nu}R$ and hence
\ba
& & \square \hat{U}=4a_2 R\,,\label{UR} \\
& & S_{\mu \nu}=\frac14 g_{\mu \nu} \hat{U}\,,
\label{Smunu2}
\ea
where $\hat{U} \equiv S^0_0+S^i_i$.
On the flat FLRW background the field $\hat{U}$ is identical
to $U$ introduced in Eq.~(\ref{Udef}), with $V=0$.
{}From Eqs.~(\ref{Smunu}) and (\ref{Smunu2}) the field equations
of motion (\ref{nonlocal2}) read
\be
G_{\mu \nu}-m^2 \left[ \frac14 g_{\mu \nu} \hat{U}
-\frac12 (\nabla_{\mu}S_{\nu}+\nabla_{\nu}S_{\mu})
\right]=8\pi GT_{\mu \nu}\,.
\label{GU}
\ee
Taking the covariant derivative of Eq.~(\ref{Smunu}), we obtain
\be
\nabla_{\nu} \hat{U}=2\left( \square S_{\nu}
+\nabla_{\mu} \nabla_{\nu}S^{\mu} \right)\,.
\label{Unu}
\ee

We decompose the field $\hat{U}$ into the background component
$U(t)$ and the perturbation $\delta U(t,{\bm x})$, as
\be
\hat{U}=U(t)+\delta U(t,{\bm x})\,.
\ee
The time component of the vector $S_{\mu}$ can be also
decomposed as $S_0=\bar{S}_0(t)+\delta S_0 (t,{\bm x})$,
where we omit the bar in the following for simplicity.
The spatial component of $S_{\mu}$ can be written
as $S_i=S_i^{\rm T}+\partial_{i} \delta S$, where
$\delta S$ is a scalar and $S_i^{\rm T}$ is a
transverse vector satisfying $\partial_i S_i^{\rm T}=0$.
Since we are interested in only scalar perturbations,
we do not consider the contribution of vector perturbations
$S_i^{\rm T}$. Then, the four-vector $S_{\mu}$
can be expressed as
\be
S_{\mu}=\left( S_0+\delta S_0,\partial_i \delta S \right)\,.
\ee

In order to derive the full perturbation equations of motion,
we need to expand Eqs.~(\ref{UR}), (\ref{GU}), and (\ref{Unu})
up to first order in perturbations.
In doing so, we consider scalar metric perturbations $\Phi$
and $\Psi$ described by the following metric in
longitudinal gauge \cite{Bardeen}:
\be
ds^2=-(1+2\Phi)dt^2+a^2(t)(1-2\Psi)\delta_{ij}dx^idx^j\,,
\label{metric}
\ee
for which the perturbations of the Ricci scalar $R$ and
the Einstein tensor $G_{\mu \nu}$, etc. can be computed.
Since our interest is the evolution of perturbations during the matter era,
we take into account a non-relativistic perfect fluid characterized
by the energy-momentum tensor:
\be
T_0^0=-(\rho_m+\delta \rho_m)\,,\qquad
T^0_i=-\rho_m \upsilon_{m,i}\,,\qquad
T^i_j=0 \, \quad (i,j=1,2,3)\,,
\ee
where $\delta \rho_m$ is the density perturbation and
$\upsilon_{m}$ is the velocity potential.

\subsection{Perturbation equations}

The perturbation $\delta T^{\mu \nu}$ of the matter
energy-momentum tensor $T^{\mu \nu}$
obeys the continuity equation,
\be
{\delta T^{\mu \nu}}_{;\mu}=0\,.
\label{Tcon}
\ee
{}From the $\nu=0$ and $\nu=i$ components of
Eq.~(\ref{Tcon}), we obtain the following equations
in Fourier space, respectively:
\ba
& &\dot{\delta \rho_m}+3H \delta \rho_m-3\rho_m \dot{\Psi}
+\frac{k^2}{a^2}\rho_m \upsilon_m=0\,,\label{per1} \\
& & \dot{\upsilon}_m=\Phi\,,\label{per2}
\ea
where $k$ is a comoving wave number.
We introduce the gauge-invariant density contrast
\be
\delta_m \equiv \frac{\delta \rho_m}{\rho_m}
+3H \upsilon_m\,.
\label{delmdef}
\ee
Taking the time derivative of Eq.~(\ref{per1}) and
using Eq.~(\ref{per2}), the density contrast satisfies
\be
\ddot{\delta}_m+2H \dot{\delta}_m+\frac{k^2}{a^2}\Phi
=3\ddot{B}+6H \dot{B}\,,
\label{delmeq}
\ee
where $B \equiv \Psi+H\upsilon_m$.

{}From the (00), $(0i)$, $(ij)$ [$i \neq j$], and the trace of
the $(ii)$ parts of the perturbed version of Eq.~(\ref{GU}),
we obtain the following equations of motion in Fourier space
respectively:
\ba
\hspace{-1cm}
& &\frac{2k^2}{a^2}\Psi+6H \left(\dot{\Psi}+H \Phi \right)
-m^2 \left( \frac14 \delta U+\dot{\delta S}_0
-2\dot{S}_0\Phi-S_0 \dot{\Phi} \right)=-8\pi G \delta \rho_m\,,
\label{per3} \\
\hspace{-1cm}
& &2\left(\dot{\Psi}+H\Phi \right)+\frac{m^2}{2}
\left( \dot{\delta S}+\delta S_0-2S_0\Phi-2H \delta S
\right)=8\pi G \rho_m \upsilon_m\,,
\label{per4} \\
\hspace{-1cm}
& &\Psi-\Phi+m^2 \delta S=0\,,
\label{per5} \\
\hspace{-1cm}
& &6 \ddot{\Psi}+6H \left( \dot{\Phi}+3 \dot{\Psi} \right)
+6 \left( 3H^2+2\dot{H} \right) \Phi-2\frac{k^2}{a^2}
(\Phi-\Psi)-m^2 \left[ \frac34 \delta U+\frac{k^2}{a^2}
\delta S+3H \delta S_0-3S_0 \left( \dot{\Psi}+2H \Phi
\right) \right]=0.
\label{per6}
\ea
{}From Eq.~(\ref{UR}) it follows that
\be
\ddot{\delta U}+3H \dot{\delta U}+\frac{k^2}{a^2}\delta U
-2\Phi \left( \ddot{U}+3H \dot{U} \right)
-\left( \dot{\Phi}+3\dot{\Psi} \right) \dot{U}
=8a_2 \left[ 3\left( \ddot{\Psi}+4H \dot{\Psi}+H \dot{\Phi}
\right)+6 \left(2H^2+\dot{H} \right)\Phi+
\frac{k^2}{a^2} \left( 2\Psi-\Phi \right) \right].
\label{per7}
\ee
The $\nu=0$ and $\nu=i$ components of Eq.~(\ref{Unu}) read
\ba
\dot{\delta U} &=&
-4 \left[ \ddot{\delta S}_0+3H \dot{\delta S}_0
-S_0 \ddot{\Phi}-2\ddot{S}_0 \Phi-3\dot{S}_0
\left( \dot{\Phi}+\dot{\Psi}+2H \Phi \right)-3H^2 \delta S_0
\right]\nonumber \\
& & +12HS_0 \left( \dot{\Phi}-2\dot{\Psi}-2H \Phi \right)
-2 \frac{k^2}{a^2} \left( \delta S_0+\dot{\delta S}
-4H \delta S-2S_0 \Phi \right)\,,
\label{per8}\\
\delta U &=& -2 \left[ \ddot{\delta S}+H \dot{\delta S}
+2\frac{k^2}{a^2} \delta S-2 \left( \dot{H}+3H^2 \right)
\delta S+\dot{\delta S}_0+5H \delta S_0
-2S_0 \left( \dot{\Phi}+\dot{\Psi}+4H \Phi \right)
-4\dot{S}_0 \Phi \right]\,.
\label{per9}
\ea

The evolution of the density contrast $\delta_m$ is known by
solving Eqs.~(\ref{per1}) and (\ref{per2}) and (\ref{per3})---(\ref{per9})
for given $k$.
In the $m \to 0$ limit, all the mass-dependent terms involving
the perturbations $\delta U$, $\delta S_0$, and $\delta S$
in Eqs.~(\ref{per3})---(\ref{per6}) vanish to recover the
general relativistic behavior.
When $m \neq 0$ the evolution of the gravitational
potentials $\Phi$ and $\Psi$ is subject to change,
which affects the growth of $\delta_m$ through
Eq.~(\ref{delmeq}).
Eliminating the terms $\dot{\Psi}+H \Phi$ from
Eqs.~(\ref{per3}) and (\ref{per4}), we obtain
\be
\frac{k^2}{a^2}\Psi-\frac{m^2}{8} \delta {\cal F}
=-4\pi G \rho_m \delta_m\,,
\label{Poi}
\ee
where
\be
\delta {\cal F} \equiv
\delta U
+4\dot{\delta S}_0+6H \delta S_0+6H \dot{\delta S}
-12H^2 \delta S-4S_0 \dot{\Phi}
-4 \left( 2\dot{S}_0+3H S_0 \right)\Phi\,.
\label{calF}
\ee
In the $m \to 0$ limit, Eq.~(\ref{Poi}) reduces to the standard Poisson
equation $(k^2/a^2)\Psi=-4\pi G \rho_m \delta_m$.
In GR we have $\delta S=0$ and $\Psi=\Phi$,
so that the third term on the left-hand side (lhs) of Eq.~(\ref{delmeq})
reads $(k^2/a^2)\Phi=-4\pi G \rho_m \delta_m$.
This term works as a driving force for the growth of $\delta_m$
with the gravitational coupling characterized by $G$.
In the presence of the mass term $m$, there is a modification
to the gravitational constant $G$.

\subsection{Evolution of perturbations on subhorizon scales}

Let us consider the perturbations relevant to the linear regime
of galaxy clusterings. This corresponds to the wave numbers
$0.01\,h$ Mpc$^{-1}$$\lesssim k \lesssim0.1\,h$ Mpc$^{-1}$ \cite{Tegmark}, i.e.,
\be
30 \lesssim k/H_0 \lesssim 300\,.
\label{modes}
\ee
In the redshift range where the redshift distortions of galaxies have been
measured ($z \equiv 1/a-1 \lesssim 2$), the modes (\ref{modes}) are
deep inside the Hubble radius ($k/a \gg H$).
Under a subhorizon approximation we can ignore some of the terms
in the perturbation equations (\ref{per3})---(\ref{per9}) (see e.g.,
Refs.~\cite{Boi,Tsujikawa:2007gd,Nesseris:2008mq}).
We also note that the gravitational potentials $\Phi$ and $\Psi$ are
nearly constant during the deep matter era and they start to vary
after the onset of the cosmic acceleration, so that
$|\dot{\Phi}| \lesssim |H \Phi|$ and $|\dot{\Psi}| \lesssim |H \Psi|$
by today.

Under the subhorizon approximation the dominant contributions
to Eq.~(\ref{per7}) should be the terms including $k^2/a^2$,
and hence
\be
\delta U \simeq 8a_2 \left( 2\Psi-\Phi \right)
=8a_2 \left( \Phi-2m^2 \delta S \right)\,,
\label{delU}
\ee
where in the second equality we used Eq.~(\ref{per5}).
For the validity of this approximation we also require that
$(k^2/a^2)|\delta U| \gg |\Phi H \dot{U}|$, which can be
interpreted as $k^2/(aH)^2 \gg |U|$ for $a_2=1/3$
and $|\dot{U}| \lesssim |HU|$.
Since today's value of $|U|$ is of the order of 10,
the condition $k^2/(aH)^2 \gg |U|$ is satisfied for the
wave numbers  (\ref{modes}).
We also note that Eq.~(\ref{per7}) does not contain
a large mass term exceeding $k/a$, so the oscillating mode
induced by the second derivative $\ddot{\delta U}$ can
be neglected relative to the mode (\ref{delU}).

We recall that the mass scale $m$ is slightly smaller than $H_0$,
in which case $k^2/a^2$ is much larger than $m^2$
for the subhorizon modes (\ref{modes}).
{}From Eqs.~(\ref{per8}) and (\ref{per9}) we can estimate
the orders of the subhorizon perturbations $\delta S$
and $\delta S_0$, as
\be
|\delta S| \approx \frac{a^2}{k^2} |\Phi|\,,\qquad
|\delta S_0| \approx \frac{a^2H}{k^2} |\Phi|\,.
\label{delS}
\ee
{}From Eq.~(\ref{per5}) it follows that
\be
\left|\frac{\Psi}{\Phi}-1 \right|=O(\epsilon_k)\,,\quad
{\rm where } \quad
\epsilon_k\equiv \frac{(ma)^2}{k^2}\,.
\label{ani}
\ee
Since $\epsilon_k \ll 1$, the difference between $\Psi$
and $\Phi$ is small.
On using Eq.~(\ref{delS}), the perturbation $\delta {\cal F}$
in Eq.~(\ref{calF}) is approximately given by
\be
\delta {\cal F} \simeq
\delta U-4S_0 \dot{\Phi}
-4 \left( 2\dot{S}_0+3H S_0 \right)\Phi\,.
\ee
The term $X=HS_0$ grows to the order close to $O(1)$ by today.
Since $\delta U=O(\Phi)$ from Eq.~(\ref{delU}),
the perturbation $|\delta {\cal F}|$ is of the order of $|\Phi|$.
From Eqs.~(\ref{Poi}) and (\ref{ani}) the third term on the lhs
of Eq.~(\ref{delmeq}) can be estimated as
\be
\frac{k^2}{a^2}\Phi \simeq -4\pi G_{\rm eff} \rho_m \delta_m\,,
\ee
where the difference between
the effective gravitational coupling $G_{\rm eff}$ and the
gravitational constant $G$ is
\be
\left|\frac{G_{\rm eff}}{G}-1 \right|=O(\epsilon_k)\,.
\label{Geff}
\ee
For the modes (\ref{modes}) the parameter $\epsilon_k$
is in the range $5 \times 10^{-6} \lesssim \epsilon_k \lesssim 5 \times 10^{-4}$,
so $G_{\rm eff}$ is very close to $G$. The rhs of Eq.~(\ref{delmeq})
can be negligible relative to its lhs, and hence
\be
\ddot{\delta}_m+2H \dot{\delta}_m
-4\pi G_{\rm eff} \rho_m \delta_m \simeq 0\,.
\label{delmsub}
\ee

Numerically we have solved the full perturbation equations (\ref{per1})---(\ref{per2})
and (\ref{per3})-(\ref{per9}) for the initial conditions
$\delta U(t_i)=\dot{\delta U}(t_i)=\delta S_0(t_i)=\dot{\delta S}_0(t_i)=
\delta S(t_i)=\dot{\delta S}(t_i)=0$.
In spite of the presence of the second derivative $\ddot{\delta U}$ in Eq.~(\ref{per7}),
the term $(k^2/a^2)\delta U$ soon starts to balance with the term $8a_2(k^2/a^2)(2\Psi-\Phi)$
on the rhs of Eq.~(\ref{per7}). After that, the solutions can be well described
by the analytic estimation given above.
Numerically we also confirmed the accuracy of Eq.~(\ref{Geff}) and found that
in practice $G_{\rm eff}/G \simeq 1$ to better than $0.05\,\%$ precision for the
wave numbers in the range (\ref{modes}).
This suggests that, apart from the difference of the background
evolution, it is difficult to distinguish the NLMG model  from the
$\Lambda$CDM model for the perturbations relevant to
large-scale structures. Therefore, in what follows and the likelihood analysis
in Sec.~\ref{resultscomp}, we shall solve Eq.~(\ref{delmsub}) together with
the background equations of motion by setting $G_{\rm eff}=G$.

\begin{figure*}[t!]
\centering
\vspace{0cm}\rotatebox{0}{\vspace{0cm}\hspace{0cm}\resizebox{0.65\textwidth}{!}{\includegraphics{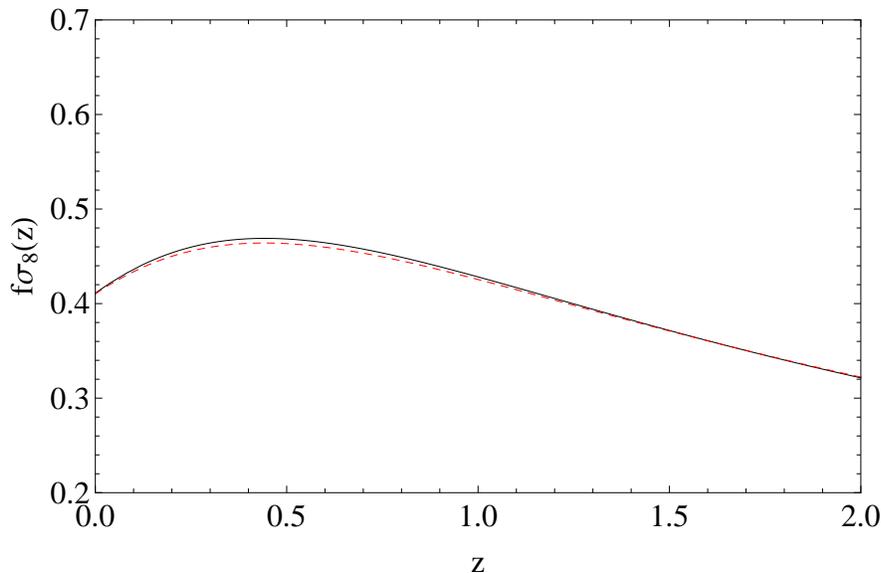}}}
\caption{The evolution of $f \sigma_8(z)$ for the NLMG model with $a_1=0$ and $a_2=1/3$,
compared to  the $\Lambda$CDM model in the redshift range $z\in[0,2]$.
The two lines correspond to $k=30H_0$ (solid black line) and the $\Lambda$CDM
model (dashed red line) for $\Omega_m^{(0)}=0.3$
and $\sigma_{8,0}=0.8$. We plot $f \sigma_8$ in this redshift range as
this is where the current data exist.
\label{fig3}}
\end{figure*}

The observations of redshift space distortions can place bounds on
the quantity $f \sigma_8$, where $f \equiv \dot{\delta}_m/(H \delta_m)$
characterizes the growth rate of matter perturbations and
$\sigma_8$ is the rms amplitude of $\delta_m$ at the comoving
$8\,h^{-1}$ Mpc scale \cite{Tegmark}.
In Fig.~\ref{fig3} we plot $f \sigma_8$ versus the redshift $z$
for the NLMG model with $a_1=0$ and $a_2=1/3$
as well as for the $\Lambda$CDM model.
Today's values of $\Omega_m$ and
$\sigma_8$ are chosen to be $\Omega_m^{(0)}=0.3$
and $\sigma_{8,0}=0.8$, respectively.
The solid black curve in Fig.~\ref{fig3} corresponds to the
wave number $k=30H_0$, but we confirmed that
the evolution of $f\sigma_8$ for $k>30H_0$ is similar
to that for $k=30H_0$.

Since $G_{\rm eff}$ is very close to $G$ for the modes (\ref{modes}),
the growth rate of matter perturbations
in the NLMG model is similar to that
in the $\Lambda$CDM model.
The main reason of the small difference seen in Fig.~\ref{fig3}
is that the background evolution of $w_{\rm DE}$ is different.
This is similar to what happens for the constant $w_{\rm DE}$
models in the framework of GR \cite{Jailson}.
For a practical purpose, the evolution of $f \sigma_8$
in the NLMG model can be known in good accuracy
by solving Eq.~(\ref{delmsub}) with $G_{\rm eff}=G$.

\section{Observational constraints\label{resultscomp}}

In this section we will confront the NLMG model (\ref{nonlocal1})
characterized by $a_1=0$ and $a_2=1/3$ with the latest cosmological
data and study whether they can be distinguished from the \lcdm model.
\subsection{The data}

In order to constrain the NLMG model, we use the same numerical
code\footnote{General minimization and MCMC cosmological codes can be
found freely available\cite{nespage}.} and
the same data of SnIa, BAO and growth rate as those
in Refs.~\cite{Basilakos:2013nfa,Nesseris:2013jea},
so we refer the readers to the aforementioned references for detail.
We also employ the correlation matrix of the Planck CMB shift parameters
($l_a, {\cal R}, z_*$) presented in Ref.~\cite{Shafer:2013pxa}.
These three parameters are related to the background
quantities such as $\Omega_m^{(0)}$, $\Omega_b^{(0)}$, and $h$.
They can efficiently summarize the CMB information on dark energy
in a model-independent way \cite{Wang}.

Compared to Refs.~\cite{Basilakos:2013nfa,Nesseris:2013jea},
there is a difference in the analysis of the growth-rate data.
Instead of using the well-known $\gamma(z)$ parametrization and modeling
the growth rate as $f(z)=\Omega_m(z)^{\gamma(z)}$, we directly fit
the numerical solution of the perturbation equations.
Regarding the data of the growth rate given in Table I of
Ref.~\cite{Basilakos:2013nfa}, they are based on the {\em WiggleZ},
SDSS, 2dF, PSCz, VVDS, 6dF, 2MASS and BOSS galaxy surveys.
The data themselves are given in terms of $f(z)\sigma_{8}(z)$.
It should be stressed that the main benefit of using $f(z)\sigma_{8}(z)$,
instead of just $f(z)$, is that the former is directly related to the power
spectrum of peculiar velocities of galaxies.

\subsection{Fitting method and model comparisons}

As mentioned in the previous section, the mass $m$ is known from
Eq.~(\ref{mH0}) and it is not a free parameter of the theory.
For the case $a_1=0$ it is known as a function of the matter density
parameter $\Omega_{m}^{(0)}$ and this is explicitly shown in the right panel
of Fig. \ref{fig2}. In the general case, regarding the value of $a_2$, it can be
fixed to 1/3 in accordance with the literature and the mass scale $m$ can as
well be computed in this case. However, its precise value does not matter
because we can always absorb the coefficient $a_2$ into $m$.

The mass scale $m$ is determined by the demand that
the system of the background equations of motion is consistent, i.e.,
the value initially used for the solution has to be the same as the one
derived from Eq.~(\ref{mH0}). Since the results depend on $\Omega_{m}^{(0)}$,
we implement an iterative algorithm in which the value of $m$ is found
for each value of $\Omega_{m}^{(0)}$ via Eq.~(\ref{mH0}) to check the
consistency of Eqs.~(\ref{auto1})---(\ref{auto6}).
These values are also saved and
used later on to simplify and speed up the fitting procedure. Therefore,
the final set of parameters employed in the minimization is
$(\Omega_{m}^{(0)}, \Omega_{b}^{(0)} h^2, \sigma_{8,0})$,
where $\Omega_b^{(0)}$ is today's baryon density parameter.
This situation is analogous to what happens in
the $\Lambda$CDM model.

We compute the total chi square
\be
\chi^2=\chi^2_{\rm SnIa}+\chi^2_{\rm BAO}
+\chi^2_{\rm CMB}+\chi^2_{\rm growth}\,,
\label{chisde}
\ee
where each term on the rhs is derived by fitting with
the SnIa, BAO, CMB, and growth-rate data, respectively,
along the line of Refs.~\cite{Basilakos:2013nfa,Nesseris:2013jea}.
The best-fit corresponds to model parameters for which
$\chi^2$ takes a minimum value $\chi^2_{\rm bf}$.
We will also consider the same parameters
$(\Omega_{m}^{(0)}, \Omega_{b}^{(0)} h^2, \sigma_{8,0})$
in the \lcdm model and evaluate the total chi square for
the comparison with the NLMG model.

Following Ref.~\cite{Shafer:2013pxa}, we discuss the effect of the $H_0$
prior on the results. The Planck team essentially pinned down the parameter
$\Omega_m^{(0)} h^2$ to very high precision, so changing $h$ also
affects $\Omega_m^{(0)}$, since $\delta(\Omega_m^{(0)} h^2) \simeq 0$
or equivalently $\delta \ln \Omega_m^{(0)} \simeq -2 \delta \ln h$.
The latter implies that, while fitting the data, increasing $h$ forces $\Omega_m^{(0)}$
to lower values and vice versa. However, due to the degeneracies in the CMB+BAO data,
a lower value of $\Omega_m^{(0)}$ implies a more negative dark energy equation of
state, i.e. $w_{\rm DE}<-1$. In simple terms, increasing $h$ reduces $\Omega_m^{(0)}$
and forces $w_{\rm DE}$ to more negative values and vice versa.
Therefore, the value $h$ that we choose is important in the rest of the analysis especially
since, as mentioned before, the NLMG model has a corresponding equation of state
$w_{\rm DE}$ between $-1.1$ and $-1.04$.
In order to accommodate the cases with different values of $H_0$, we will
test some priors on $h$: (i) the Planck best fit: $h=0.673$ \cite{Ade:2013zuv},
(ii) the best-fit $h=0.738$ derived by the direct measurement of
$H_0$ \cite{Riess:2011yx}, and (iii) other four values of $h$ ranging
$0.673<h<0.738$.

\begin{table*}[!]
\tabcolsep 4.5pt
\vspace{1mm}
\begin{tabular}{ccccccc} \hline \hline
Model     & $\Omega_{m}^{(0)}$     & $\Omega_{b}^{(0)} h^2$  & $\sigma_{8,0}$&
$\chi_{\rm bf}^{2}$ &${\rm AIC}$& $|\Delta$AIC$|$ \vspace{0.05cm}\\ \hline
$h=0.673$ &&&&&& \\
$\Lambda$CDM & $0.328\pm0.002$ & $0.0234\pm 0.0002$& $0.735\pm0.019$
& 583.470& 589.470& 0 \vspace{0.01cm}\\
NLMG   & $0.334\pm0.002$ & $0.0223\pm 0.0002$&
$0.726\pm 0.019$& 585.570& 591.570& 2.100\vspace{0.01cm}\\
\hline
$h=0.738$ &&&&&& \\
$\Lambda$CDM & $0.252\pm0.002$ & $0.0249\pm 0.0002$& $0.789\pm0.021$
& 599.620& 605.620& 10.011 \vspace{0.01cm}\\
NLMG  & $0.257\pm0.002$ & $0.0245\pm 0.0002$&
$0.775\pm 0.020$& 589.609& 595.609& 0\vspace{0.01cm}\\
\hline\hline
\end{tabular}
\caption[]{Statistical results of the overall likelihood analysis:
The first column indicates the model, while the second, third, and fourth
columns provide the $\Omega_{m}^{(0)}$, $\Omega_{b}^{(0)} h^2$,
and $\sigma_{8,0}$ best-fit values.
The last three columns present the goodness-of-fit statistics
($\chi^{2}_{\rm bf}$, AIC and $\Delta {\rm AIC}_{i,j}={\rm AIC}_i-{\rm AIC}_j$).
All the error estimates come from the inverse of the Fisher matrix.
The upper part of Table shows the values for the Planck prior
$h=0.673$ \cite{Ade:2013zuv},
while the lower half the corresponding values for the Riess {\it et al.}
prior $h=0.738$ \cite{Riess:2011yx}. \label{tab:results}}
\end{table*}

%
\begin{figure*}[!t]
\centering
\vspace{0cm}\rotatebox{0}{\vspace{0cm}\hspace{0cm}\resizebox{0.47\textwidth}{!}{\includegraphics{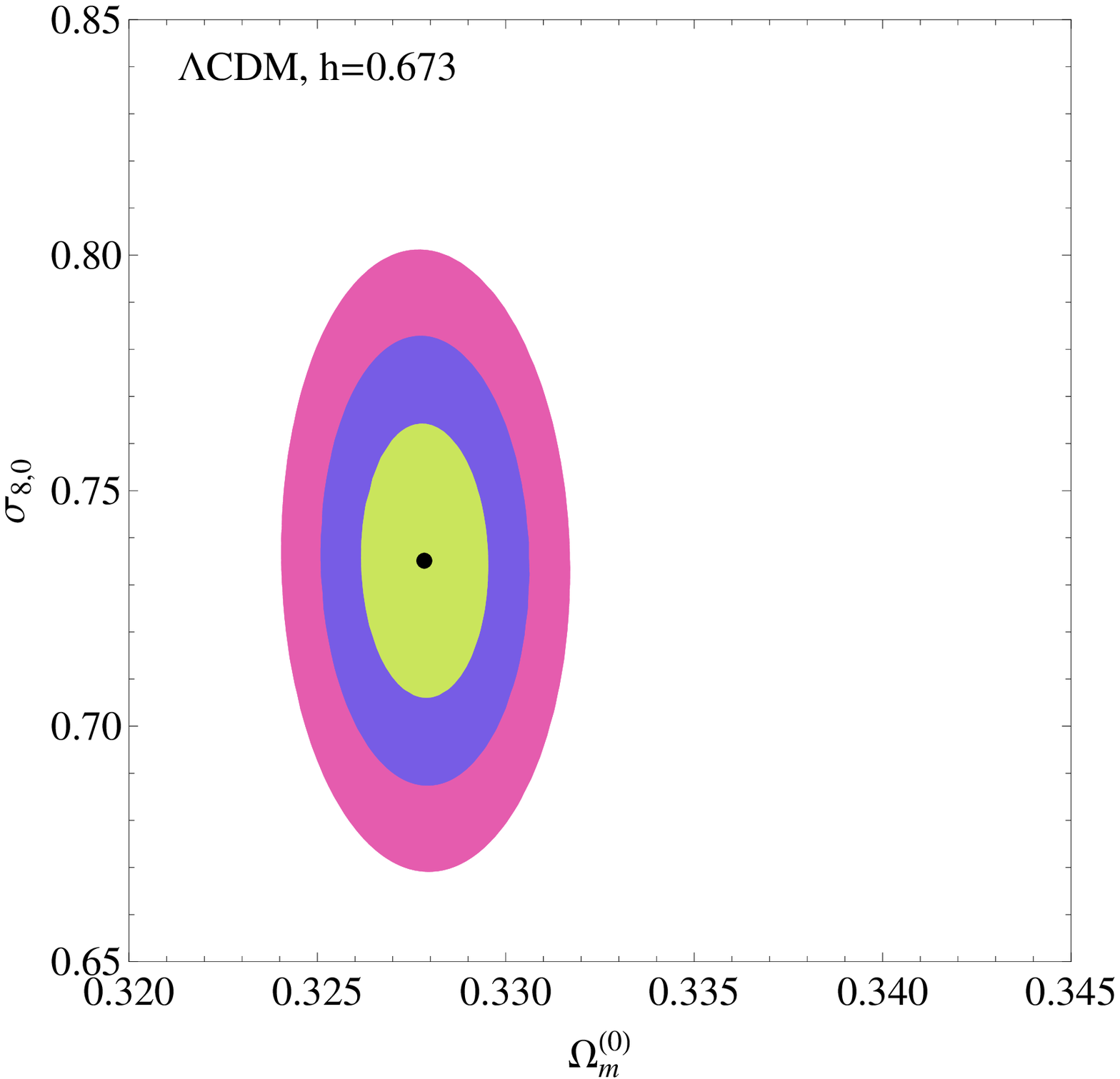}}}
\vspace{0cm}\rotatebox{0}{\vspace{0cm}\hspace{0cm}\resizebox{0.47\textwidth}{!}{\includegraphics{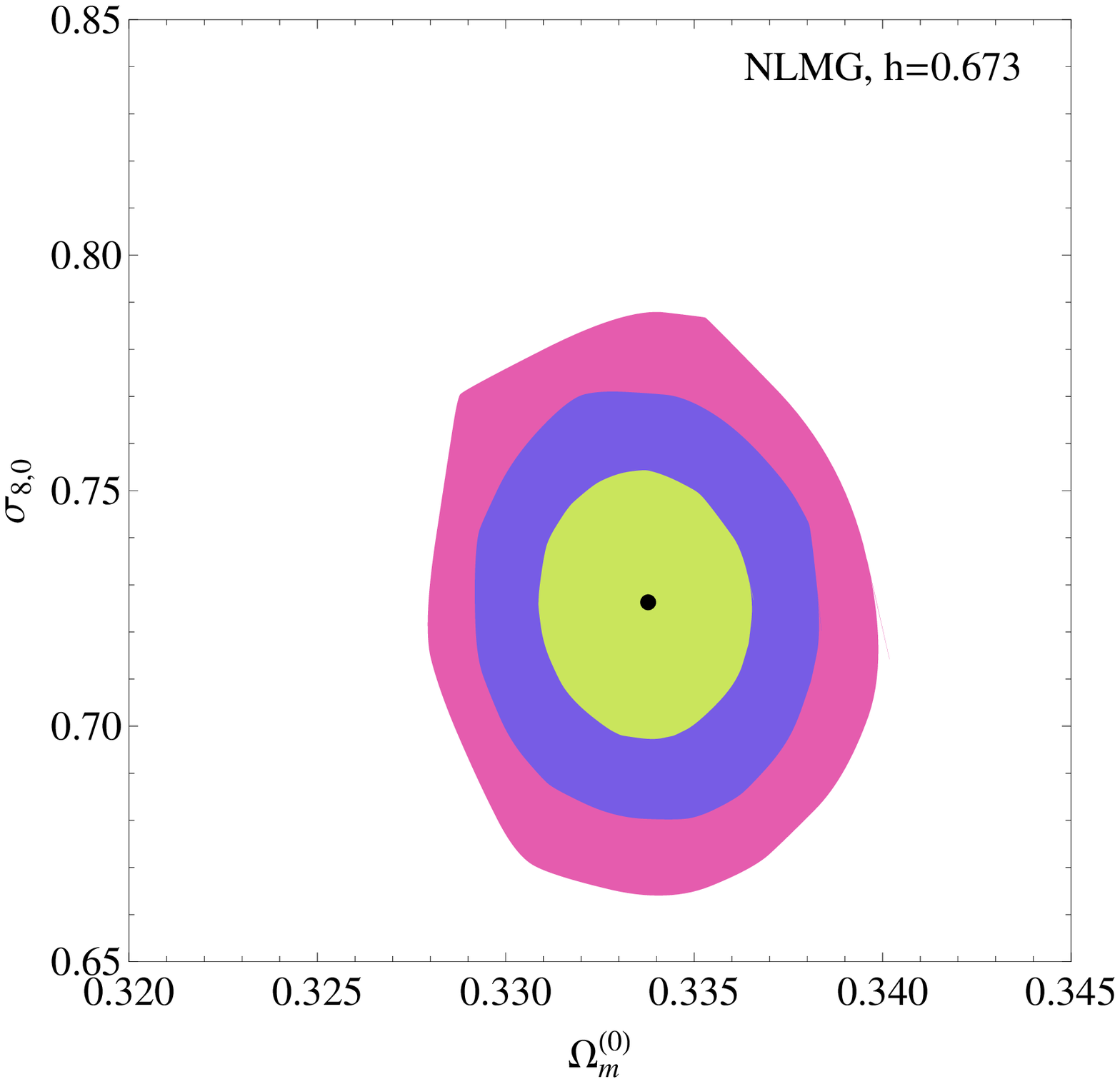}}}
\vspace{0cm}\rotatebox{0}{\vspace{0cm}\hspace{0cm}\resizebox{0.47\textwidth}{!}{\includegraphics{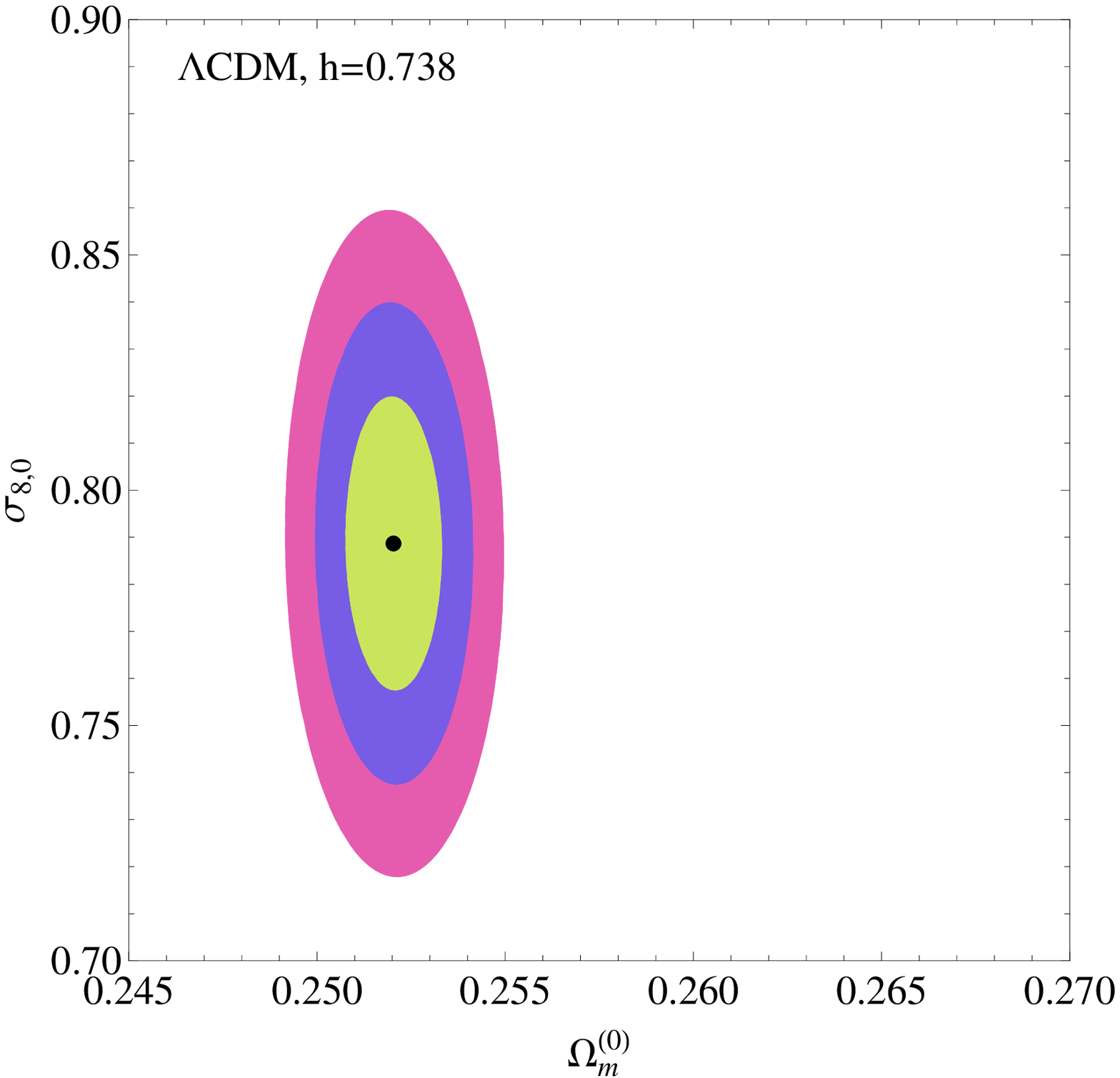}}}
\vspace{0cm}\rotatebox{0}{\vspace{0cm}\hspace{0cm}\resizebox{0.47\textwidth}{!}{\includegraphics{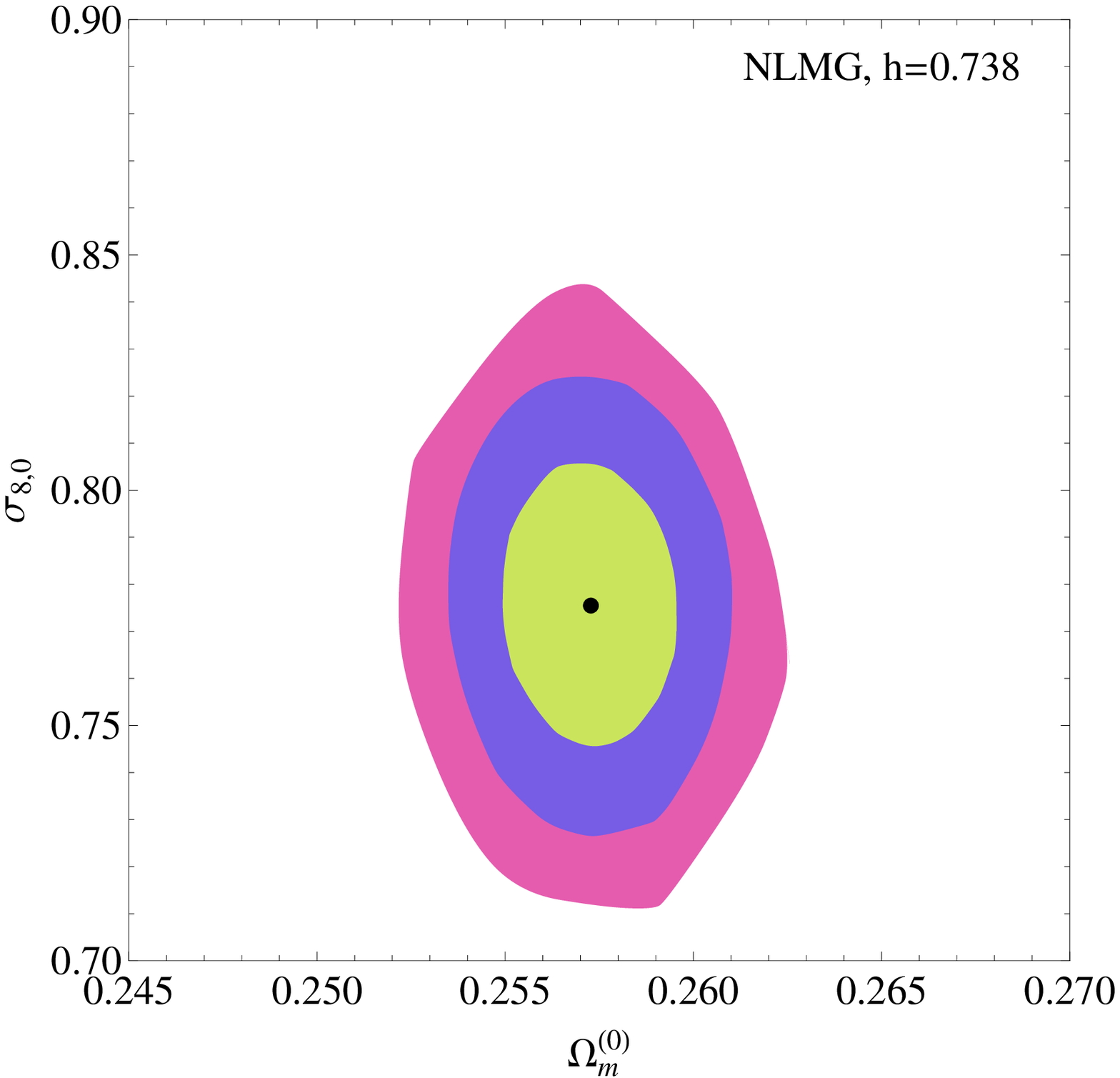}}}
\caption{The 1, 2 and 3$\sigma$ contour plots for the \lcdm (left) and the NLMG (right)
for $a_1=0$ and $a_2=1/3$ in the $(\Omega_m^{(0)},\sigma_{8,0})$ parameter space.
The top row shows the contours for $h=0.673$ and the bottom row
for $h=0.738$. \label{oms8plot}}
\end{figure*}

We will also consider the Akaike information criterion (AIC) \cite{Akaike1974}
as in Refs.~\cite{Basilakos:2013nfa,Nesseris:2013jea}.
The AIC is defined, for the case of Gaussian errors, as
\be
{\rm AIC}=\chi^2_{\rm bf}+2\ell\,,
\ee
where $\ell$ is the number of free parameters.
A smaller value of the AIC indicates a better fit to the data.
In order to effectively compare two different models, we need to estimate
the differences $\Delta {\rm AIC}_{1,2} = {\rm AIC}_{1} - {\rm AIC}_{2}$ for the two
models $1$ and $2$.
Since $\ell=3$ in both the NLMG and the $\Lambda$CDM models,
the difference of AIC between the two models is actually equivalent
to that of $\chi^2_{\rm bf}$ between them.
The larger the value of $|\Delta{\rm AIC}|$,
the higher the evidence against the model with a larger value of ${\rm AIC}$,
with a difference $|\Delta {\rm AIC}|\, \magcir 2$ indicating weak evidence
and $|\Delta {\rm AIC}|\, \magcir 6$ indicating a stronger evidence in favor of
the model with smaller AIC, while a value $\mincir 2$ indicates consistency
among the two comparison models. The AIC penalizes models with more parameters,
but it should be stressed that these numbers are provided only
as a rule of thumb and they should be used with caution \cite{Nesseris:2012cq}.

\subsection{Results}

\begin{figure*}[!t]
\centering
\vspace{0cm}\rotatebox{0}{\vspace{0cm}\hspace{0cm}\resizebox{0.47\textwidth}{!}{\includegraphics{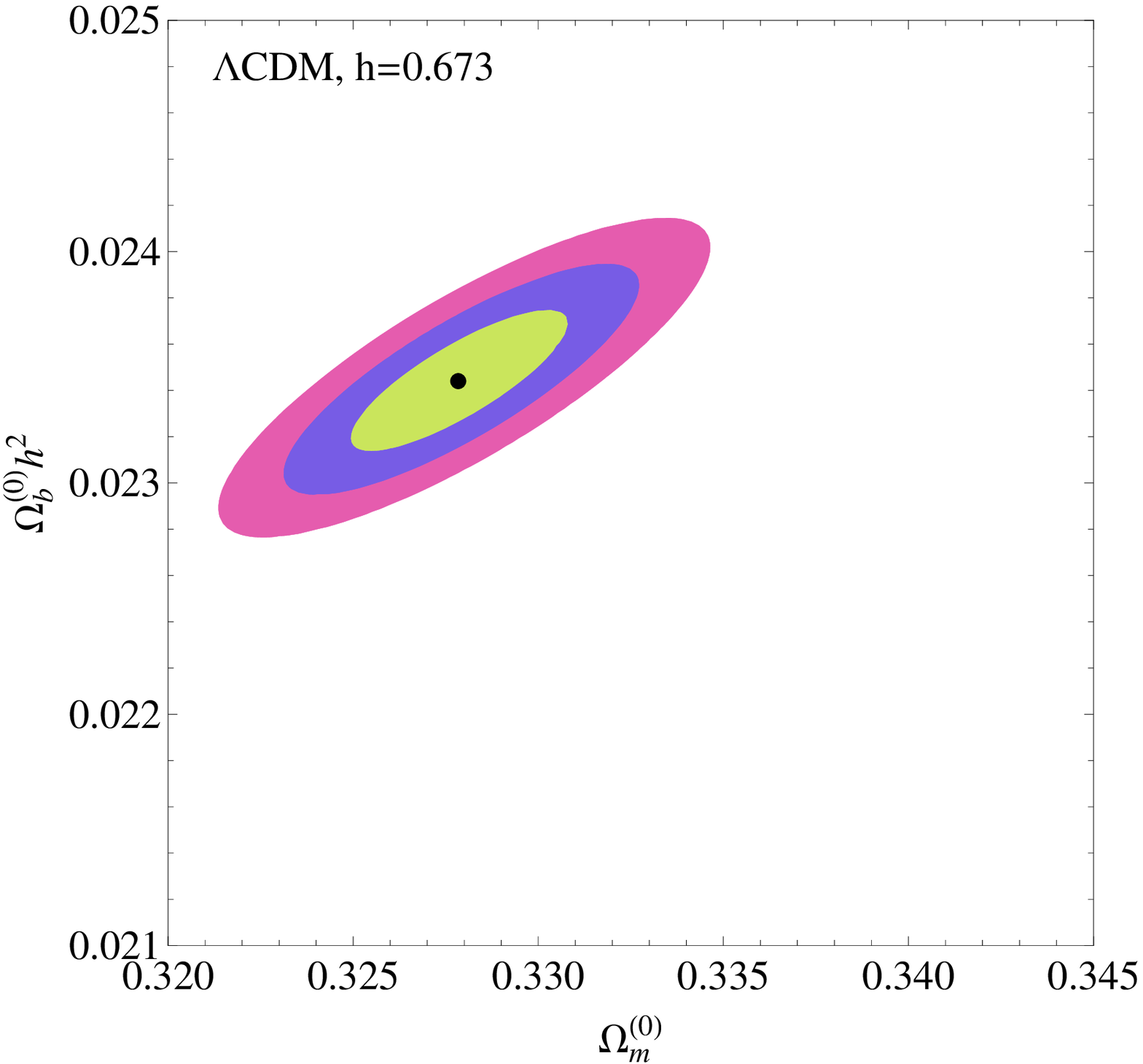}}}
\vspace{0cm}\rotatebox{0}{\vspace{0cm}\hspace{0cm}\resizebox{0.47\textwidth}{!}{\includegraphics{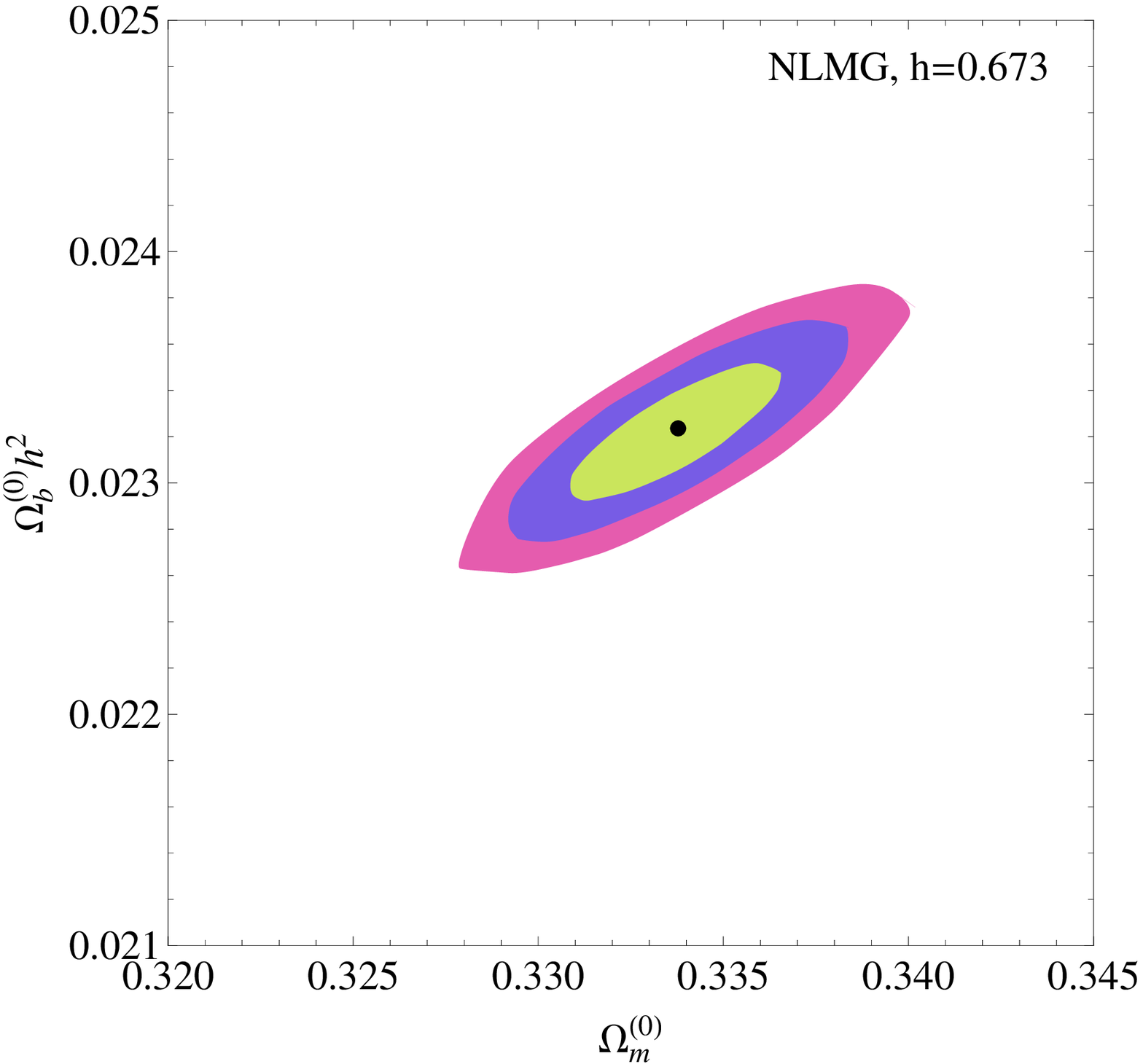}}}
\vspace{0cm}\rotatebox{0}{\vspace{0cm}\hspace{0cm}\resizebox{0.47\textwidth}{!}{\includegraphics{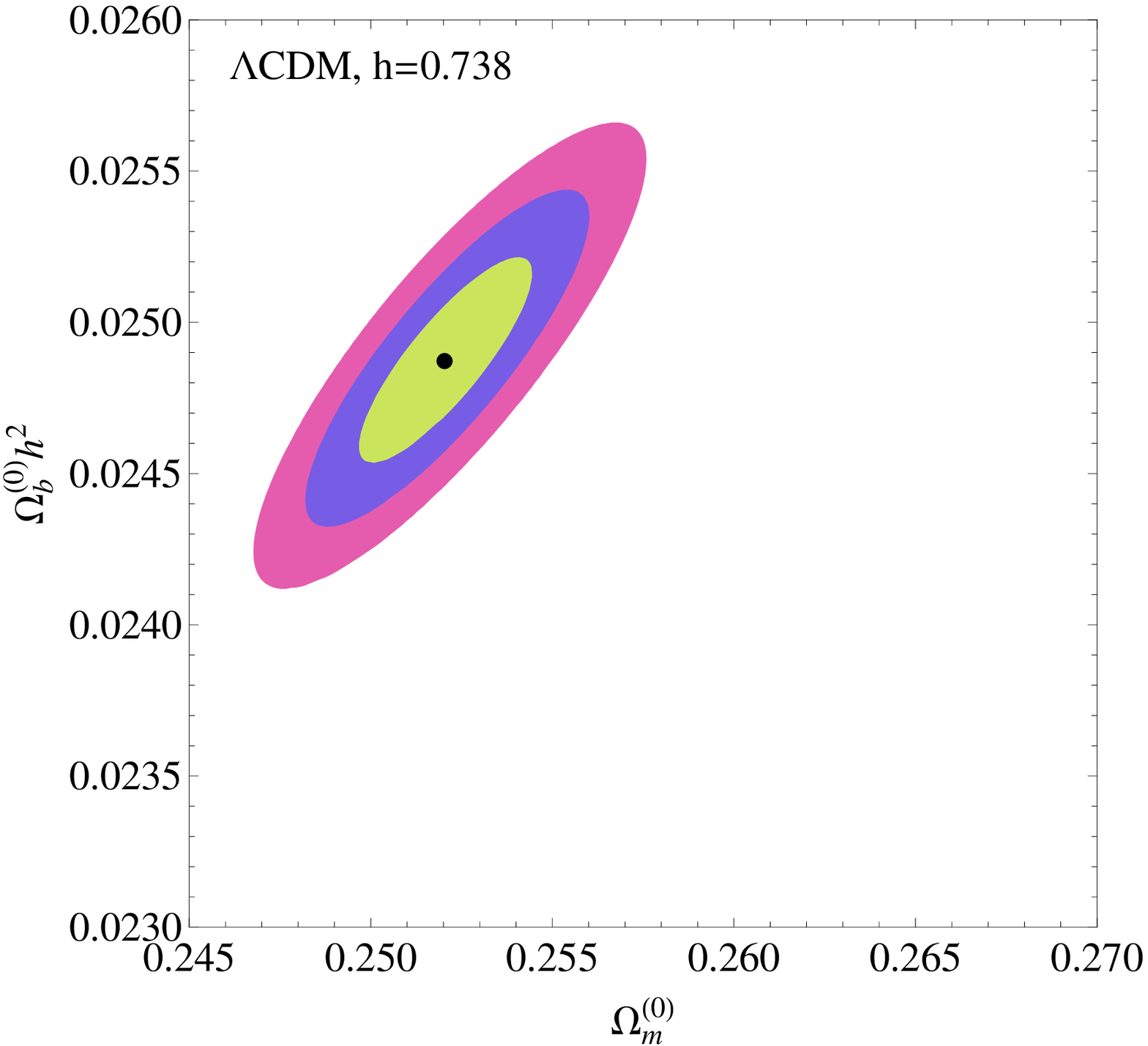}}}
\vspace{0cm}\rotatebox{0}{\vspace{0cm}\hspace{0cm}\resizebox{0.47\textwidth}{!}{\includegraphics{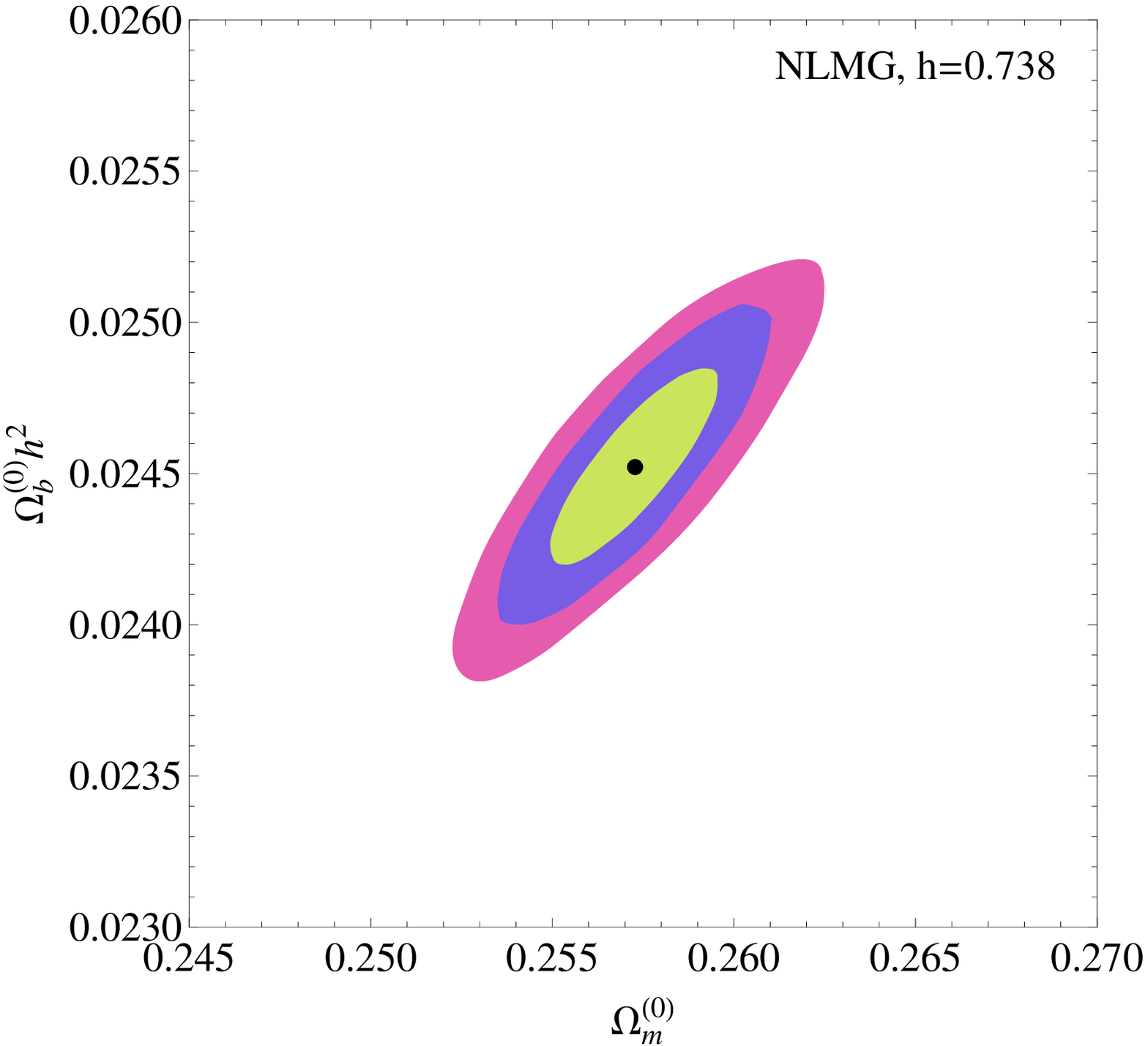}}}
\caption{The 1, 2 and 3$\sigma$ contour plots for the \lcdm (left) and the NLMG (right) for $a_1=0$ and $a_2=1/3$
in the $(\Omega_m^{(0)},\Omega_b^{(0)} h^2)$ parameter space.
The top row shows the contours for $h=0.673$ and the bottom row for $h=0.738$.\label{omobh2plot}}
\end{figure*}

In Table \ref{tab:results} we show the best-fit values of
$\Omega_{m}^{(0)}$, $\Omega_{b}^{(0)} h^2$, and
$\sigma_{8,0}$ both for the $\Lambda$CDM
and the NLMG models with $a_1=0$ and $a_2=1/3$
with two different values of $h$.
In Figs.~\ref{oms8plot} and \ref{omobh2plot} we also plot the
1, 2 and 3$\sigma$ observational
contours in the ($\Omega_m^{(0)}, \sigma_{8,0}$) and
($\Omega_m^{(0)}, \Omega_b^{(0)}h^2$) planes, respectively,
for the two models with $h=0.673$ and $h=0.738$.
For the Planck prior $h=0.673$ the minimum chi
square in the NLMG is found to be
$\chi^{2}_{\rm bf}=585.570$,
which is slightly larger than that in the $\Lambda$CDM
($\chi^{2}_{\rm bf}=583.470$).
Since the difference of AIC between the two models is
$|\Delta$AIC$|=2.100$, either of them is not particularly
favored over the other.

For $h=0.673$, if we divide the minimum chi square
as Eq.~(\ref{chisde}), each contribution is given by
$\chi^2_{\rm bf,SnIa}=566.478$,
$\chi^2_{\rm bf,BAO}=7.84942$,
$\chi^2_{\rm bf,CMB}=3.67004$,
$\chi^2_{\rm bf,growth}=7.57215$
in the NLMG and
$\chi^2_{\rm bf,SnIa}=568.399$,
$\chi^2_{\rm bf,BAO}=6.62794$,
$\chi^2_{\rm bf,CMB}=0.829078$,
$\chi^2_{\rm bf,growth}=7.61382$
in the $\Lambda$CDM.
The growth data do not provide any significant
difference between the two models as expected,
whereas the SnIa data alone prefer the NLMG
to the $\Lambda$CDM.
The $\chi^2_{\rm bf,CMB}$ in the NLMG is larger than
that in the $\Lambda$CDM with the large difference $2.842$,
which is the main reason why the total $\chi^2_{\rm bf}$
in the former exceeds that in the latter for $h=0.673$.

\begin{figure*}[!t]
\centering
\vspace{0cm}\rotatebox{0}{\vspace{0cm}\hspace{0cm}\resizebox{0.47\textwidth}{!}{\includegraphics{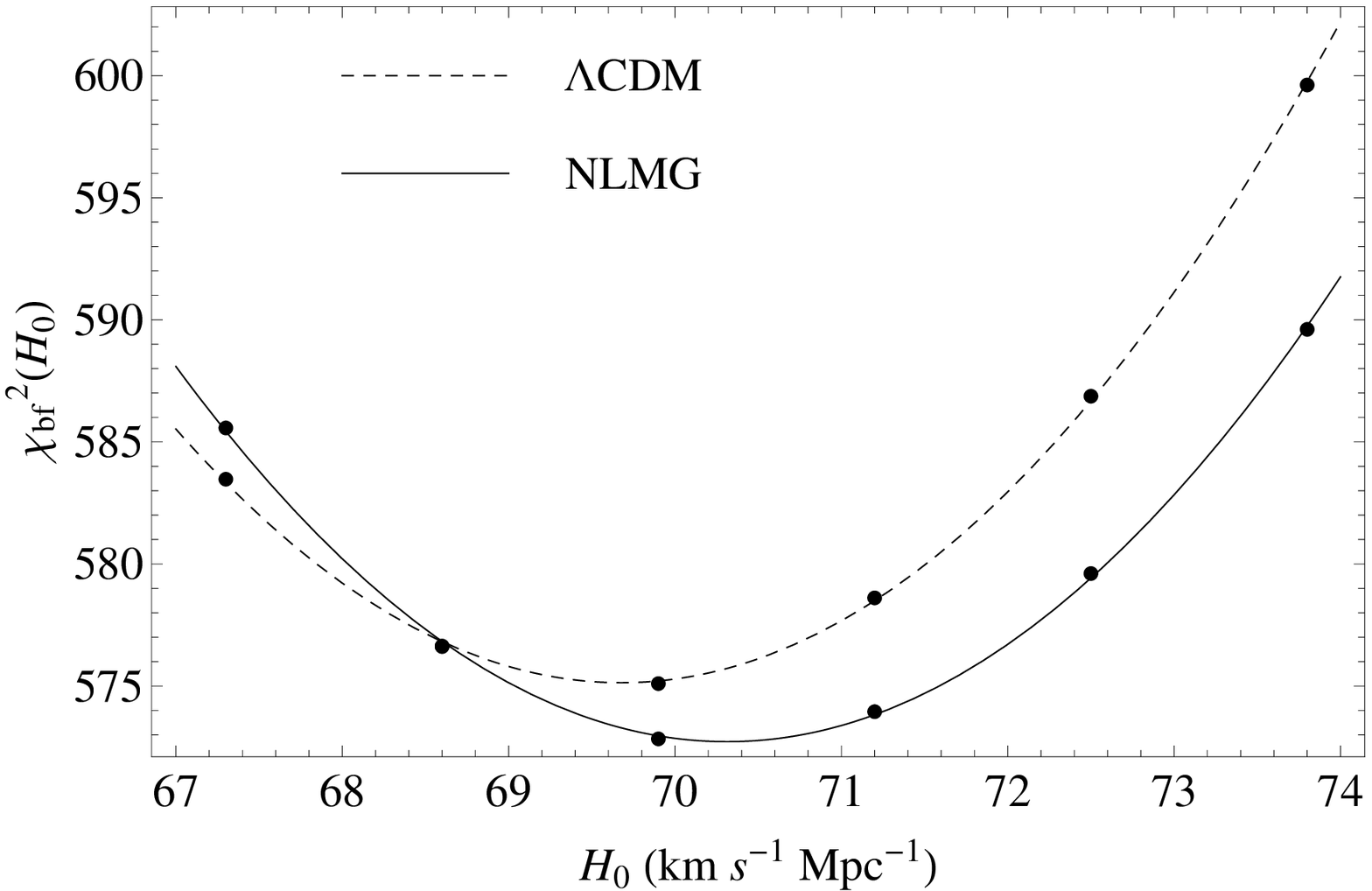}}}
\vspace{0cm}\rotatebox{0}{\vspace{0cm}\hspace{0cm}\resizebox{0.47\textwidth}{!}{\includegraphics{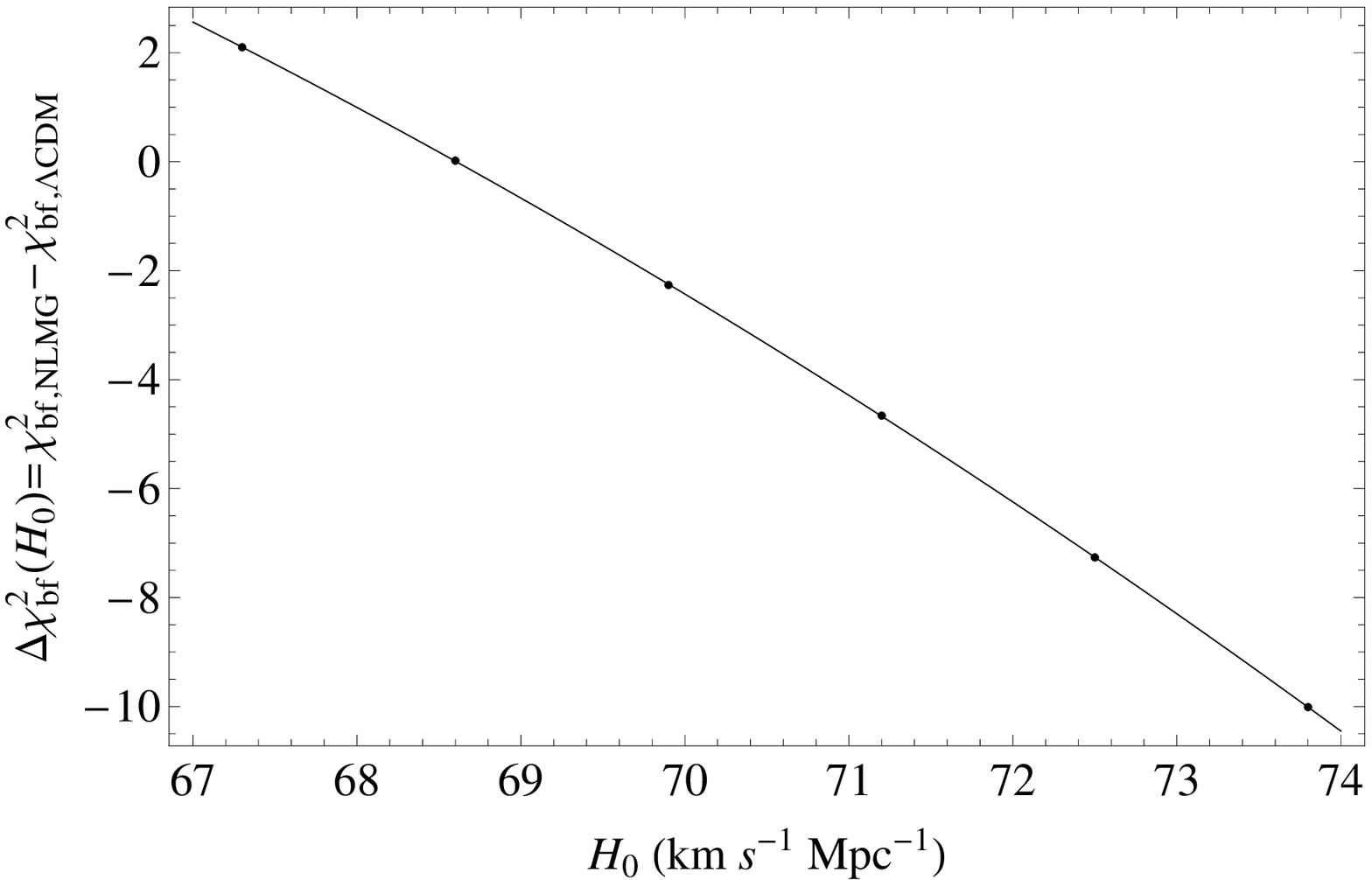}}}
\caption{Left: The value of the best-fit $\chi^2$ as a function of the prior $H_0$ for
both the NLMG (solid line) and the \lcdm (dashed line) models, respectively.
Right: The difference in the best-fit models between the NLMG and the $\Lambda$CDM,
fitted by a quadratic function (see the text). Clearly, higher values of the $H_0$
prior strongly prefer the NLMG compared
to the $\Lambda$CDM. \label{dchi2}}
\end{figure*}

For $h=0.738$ the best-fit NLMG and $\Lambda$CDM models
correspond to $\chi^2_{\rm bf}=589.609$ and
$\chi^2_{\rm bf}=599.620$, respectively.
Hence, the NLMG is significantly favored
over the $\Lambda$CDM with the difference
$|\Delta$AIC$|=10.011$.
In this case, the contributions to the minimum chi
square are
$\chi^2_{\rm bf,SnIa}=565.929$,
$\chi^2_{\rm bf,BAO}=5.38663$,
$\chi^2_{\rm bf,CMB}=10.3146$,
$\chi^2_{\rm bf,growth}=7.97966$
in the NLMG and
$\chi^2_{\rm bf,SnIa}=564.010$,
$\chi^2_{\rm bf,BAO}=6.96224$,
$\chi^2_{\rm bf,CMB}=20.5558$,
$\chi^2_{\rm bf,growth}=8.08938$
in the $\Lambda$CDM.
Therefore, both the CMB and BAO data
prefer the NLMG to the $\Lambda$CDM.
For larger $h$, the constrained regions
in Figs.~\ref{oms8plot} and \ref{omobh2plot}
shift toward smaller  values of $\Omega_m^{(0)}$.
For $\Omega_m^{(0)}$ around $0.25 \sim 0.26$, the models
with $w_{\rm DE}<-1$ are favored over the $\Lambda$CDM.
We note that the growth data do not provide any significant
difference between the two models.
The change of the bounds on $\sigma_{8,0}$ relative to the
case $h=0.673$ (seen in Fig.~\ref{oms8plot}) mainly
comes from the shift of $\Omega_m^{(0)}$.

We also perform a more extensive analysis with intermediate values of
the $H_0$ prior as well. In Fig.~\ref{dchi2} we show on the left panel the values
of the best-fit $\chi^2$ as a function of the prior $H_0$ for both the NLMG (solid line)
and the \lcdm (dashed line) models, respectively.
On the right panel we plot the difference of $\chi_{\rm bf}^2$
between the NLMG and the \lcdm models.
As a function of $h$, $\chi_{\rm bf}^2$ has a minimum around
$h_1=0.703$ in the NLMG and $h_2=0.697$ in the $\Lambda$CDM.
We adopt the following quadratic functions expanded around
$h_1$ and $h_2$ respectively:
\ba
\chi^2_{\rm bf,NLMG}(h)&=&\chi^2_{\rm bf,NLMG}(h_1)
+\frac{1}{2}\partial^2_h \chi^2_{\rm bf,NLMG}(h_1)(h-h_1)^2+ \ldots, \nn\\
\chi^2_{{\rm bf},\Lambda\textrm{CDM}}(h)&=&\chi^2_{{\rm bf},\Lambda\textrm{CDM}}(h_2)
+\frac{1}{2}\partial^2_h\chi^2_{{\rm bf},\Lambda\textrm{CDM}}(h_2)(h-h_2)^2+ \ldots ,
\label{diffchi2eqs}
\ea
where we used the fact that at the minimum the first derivative is zero.
Taking the difference, we obtain
\be
\Delta \chi^2_{\rm bf}(h)\equiv \chi^2_{\rm bf,NLMG}(h)
-\chi^2_{{\rm bf},\Lambda{\rm CDM}}(h)
= b_1+b_2 h+b_3 h^2+ \ldots,
\ee
where the constants $(b_1,b_2,b_3)$ are related to the various terms of
Eqs.~(\ref{diffchi2eqs}).
As we see in Fig.~\ref{dchi2}, this fit shows good agreement with $\chi_{\rm bf}^2$
derived for some discrete values of $h$.
For $h>0.686$, $\chi_{\rm bf}^2$ in the NLMG is smaller than that
in the $\Lambda$CDM.
In particular, for $h>0.70$, the NLMG is favored over the
$\Lambda$CDM according to the AIC.

When we consider the general case with a small but nonvanishing $a_1$,
the instability of the field $V$ leads to the dark energy equation of state
much smaller than $-1$. We have carried out the joint data analysis based on the
SnIa, CMB, and BAO data for the NLMG model with
$a_1=0.01$ and $a_2=1/3$ and for the $\Lambda$CDM model.
For $h=0.738$ we find that the bestfits correspond to $\chi_{\rm bf}^2=727.802$
in the NLMG and $\chi_{\rm bf}^2=591.528$ in the $\Lambda$CDM, respectively.
The difference of AIC from the model $a_1=0$ and $a_2=1/3$ is
$|\Delta$AIC$|\sim140$, so the models with $a_1 \neq 0$ are significantly
disfavored from the data.

\section{Conclusions}
\label{consec}

In this paper we studied cosmological perturbations and observational
constraints on the NLMG model.
Our analysis of the background cosmology covers the two models
given by Eqs.~(\ref{nonlocal0}) and (\ref{nonlocalmo}).
We dealt with (\ref{nonlocal1}) as an effective classical
equation of motion for discussing the cosmology
relevant to dark energy.
The issues of ghosts and ultraviolet completion should be addressed
in a more fundamental theory with a Lagrangian implementing quantum
and classical averaging.

We derived the background equations of motion from (\ref{nonlocal1})
on the flat FLRW background.
For the models with $a_1 \neq 0$ there is an instability for the field $V$
induced by the $-8V$ term on the lhs of Eq.~(\ref{auto2}).
In this case the mass $m$ is required to be much smaller than
$H_0$ to avoid the early onset of the cosmic acceleration.
Since the dark energy equation of state significantly deviates from $-1$,
the models with $a_1 \neq 0$ are strongly disfavored from
the joint analysis of the SnIa, CMB, and BAO data.

For the models with $a_1=0$ the rhs of Eq.~(\ref{auto2}) vanishes, so that
the field $V$ does not grow for the appropriate initial conditions (\ref{initial}).
In order for the dark energy density $\rho_{\rm DE}$ to be positive, we found
that the parameter $a_2$ has to satisfy the condition $a_2>0$.
The dark energy equation of state $w_{\rm DE}$ in the deep matter
era can be estimated as Eq.~(\ref{wdeana}), which shows good agreement
with the numerically integrated solution ($-1.1<w_{\rm DE}<-1.04$
for $-4<\ln a<0$).

Expanding the field equations of motion and the metric to first order in perturbations
about the flat FLRW background, we derived the full equations of
cosmological perturbations for the NLMG model (\ref{nonlocalmo}).
The behavior of perturbations is also estimated
for the modes relevant to galaxy clusterings.
We found that the effective gravitational coupling $G_{\rm eff}$
is very close to the gravitational constant $G$ for subhorizon
perturbations characterized by the wave numbers (\ref{modes}).
Therefore, the evolution of $f\sigma_8$ is similar to that in the
$\Lambda$CDM model (see Fig.~\ref{fig3}).
In this sense, the current growth-rate measurement alone is not able to
distinguish between the NLMG and the $\Lambda$CDM models.

We compared the NLMG model (\ref{nonlocalmo}) against the latest
cosmological observations, including SnIa, BAO, CMB, and redshift
space distortions. Since the mass $m$ is not a free parameter,
we developed an iterative algorithm to compute $m$ for
each value of $\Omega_{m}^{(0)}$ via Eq.~(\ref{mH0}) and
checked the consistency of Eqs.~(\ref{auto1})---(\ref{auto6}).
The mass $m$ in the NLMG plays a similar role to the cosmological
constant $\Lambda$ in the $\Lambda$CDM,
such that the number of free parameters in the two models is the same.
For the sake of comparison we considered the same parameters
$(\Omega_m^{(0)}, \Omega_b^{(0)} h^2, \sigma_{8,0})$ in both models.

The observational constraints on the NLMG model mainly come from
the background expansion history rather than the growth history.
The dark energy equation of state varies slowly from the deep matter
era to today around $-1.1 \lesssim w_{\rm DE} \lesssim -1.04$.
Since $w_{\rm DE}$ is approximately constant in the past,
the situation is quite similar to the case of the constant
$w_{\rm DE}$ models studied in Ref.~\cite{Shafer:2013pxa}.
The likelihood results depend on the value of the $H_0$ prior due
to the degeneracies of the CMB parameters.

We computed the chi squares in both the NLMG and the $\Lambda$CDM
models for several different values of $H_0$ ranging from the Planck
best fit $h=0.673$ \cite{Ade:2013zuv} to the Riess {\it et al.} best fit
$h=0.738$ \cite{Riess:2011yx}.
The results of our analysis are presented in Table \ref{tab:results}
and Figs.\,\ref{oms8plot}-\ref{dchi2}.
For $0.67 \lesssim h \lesssim 0.70$ the AIC shows that the NLMG and
the $\Lambda$CDM models are statistically comparable, but for
$h \gtrsim 0.70$ the NLMG model is strongly favored over the
$\Lambda$CDM model.
We hope that future observations will pin down the values of
$h$ to exquisite accuracy, clarifying whether
the NLMG model is really preferred to the models
with $w_{\rm DE} \ge -1$.


\section*{Acknowledgements}

The authors would like to thank Domenico Sapone for useful discussions.
S.T. thanks Gianluca Calcagni for the invitation to Instituto de Estructura
de la Materia (CSIC) at which this work was initiated.
S.N. acknowledges financial support from the Madrid Regional
Government (CAM) under Program No. HEPHACOS S2009/ESP-1473-02, from MICINN
under Grant No. AYA2009-13936-C06-06 and Consolider-Ingenio 2010 PAU
(CSD2007-00060), as well as from the European Union Marie Curie Initial
Training Network UNILHC PITN-GA-2009-237920. S.N. also acknowledges the support
of the Spanish MINECO's ``Centro de Excelencia Severo Ochoa" program
under Grant No. SEV-2012-0249. S.T. is supported by the Scientific Research Fund of the
JSPS (No.~24540286) and Scientific Research on Innovative Areas (No.~21111006).

\end{document}